\documentclass[lettersize,transaction]{IEEEtran}
\usepackage{enumitem,kantlipsum}

\newlist{methods}{enumerate}{1}
\setlist[methods]{%
  label={\arabic*.},
  ref={\arabic*},
  wide,
  listparindent=-\labelwidth
}
\NewDocumentCommand{\newmethod}{ m }{\item \textbf{#1}}

\usepackage{hyperref}
\usepackage{pifont}
\usepackage{listings}
\usepackage{xcolor} 
\usepackage{svg}
\usepackage{mathtools, cuted}
\usepackage{amsmath} 
\usepackage{textcomp}
\lstdefinestyle{arduinoStyle}{
  language=C++,                
  backgroundcolor=\color{white}, 
  basicstyle=\ttfamily\footnotesize, 
  keywordstyle=\color{blue},     
  commentstyle=\color{green},    
  stringstyle=\color{red},       
  numbers=left,                 
  numberstyle=\tiny\color{gray}, 
  stepnumber=1,                  
  numbersep=5pt,                 
  tabsize=2,                     
  showspaces=false,              
  showstringspaces=false,        
  showtabs=false,                
  breaklines=true,               
  breakatwhitespace=false,       
  frame=single,                  
  captionpos=b,                  
}

\IEEEoverridecommandlockouts
\usepackage{cite}
\usepackage{amssymb,amsfonts}
\usepackage{algorithmic}
\usepackage{graphicx}
\usepackage{textcomp}
\usepackage{epstopdf} 
\usepackage{epsfig}
\usepackage{subfigure}
\usepackage[super]{nth}
\usepackage{subcaption}
\usepackage{soul}
\usepackage{tabularray}
\usepackage{tabularx}
\usepackage{array}

\usepackage{graphics}
\usepackage{multirow}
\usepackage{hhline}
\DeclareUnicodeCharacter{2009}{\,} 
{\color{blue}}%
{}

\def\BibTeX{{\rm B\kern-.05em{\sc i\kern-.025em b}\kern-.08em
    T\kern-.1667em\lower.7ex\hbox{E}\kern-.125emX}}
\begin{document}

\title{An Energy-Aware RIoT System: Analysis, Modeling and Prediction in the SUPERIOT Framework}

\author{
    \IEEEauthorblockN{
    Mohammud J. Bocus\IEEEauthorrefmark{2},
    Juha H{\"{a}}kkinen\IEEEauthorrefmark{3},
    H{\'{e}}lder Fontes\IEEEauthorrefmark{5},
    Marcin Drzewiecki\IEEEauthorrefmark{4},
    Senhui Qiu\IEEEauthorrefmark{2},\\
    Kerstin Eder\IEEEauthorrefmark{2},    
    Robert Piechocki\IEEEauthorrefmark{2}    
    }    \\ 
    \IEEEauthorblockA{\IEEEauthorrefmark{2} School of Computer Science, University of Bristol, Bristol, UK}\\   
    \IEEEauthorblockA{\IEEEauthorrefmark{3} Circuits and Systems Research Unit, University of Oulu, Oulu, Finland}\\   
    \IEEEauthorblockA{\IEEEauthorrefmark{4} MpicoSys Embedded Pico Systems Sp. z o. o., Gdynia, Poland and Faculty of Electrical and Control Engineering, Gda\'{n}sk University of Technology, Gda\'{n}sk, Poland}\\ 
 \IEEEauthorblockA{\IEEEauthorrefmark{5} 
 INESC TEC, Porto, Portugal, and Faculty of Engineering, University of Porto, Porto, Portugal} \\
    \{junaid.bocus, senhui.qiu, kerstin.eder,  r.j.piechocki\}@bristol.ac.uk, juha.hakkinen@oulu.fi, helder.m.fontes@inesctec.pt,
    marcin.drzewiecki@mpicosys.com
}

\maketitle
\begin{abstract}
This paper presents a comprehensive analysis of the energy consumption characteristics of a Silicon (Si)-based Reconfigurable IoT (RIoT) node developed in the initial phase of the SUPERIOT project, focusing on key operating states, including Bluetooth Low Energy (BLE) communication, Narrow-Band Visible Light Communication (NBVLC), sensing, and E-ink display. Extensive measurements were conducted to establish a detailed energy profile, which serves as a benchmark for evaluating the effectiveness of subsequent optimizations and future node iterations. To minimize the energy consumption, multiple optimizations were implemented at both the software and hardware levels, achieving a reduction of over 60\% in total energy usage through software modifications alone. 
Further improvements were realized by optimizing the E-ink display driving waveform and implementing a very low-power mode for non-communication activities.
Based on the measured data, three measurement-based energy consumption models were developed to characterize the energy behavior of the node under: (i) normal, unoptimized operation, (ii) low-power, software-optimized operation, and (iii) very low-power, hardware-optimized operation. These models, validated with new measurement data, achieved an accuracy exceeding 97\%, confirming their reliability for predicting energy consumption in diverse configurations. 
\end{abstract}

\vspace{-3mm}
\section{Introduction}
The rapid proliferation of Internet of Things (IoT) devices has led to a wide range of applications, including smart cities, smart industrial automation and agriculture, E-Health, and environmental sensing \cite{ATZORI20102787,7879243}. The deployment of such devices in large-scale networks is often constrained by limited power availability, necessitating efficient power management strategies to ensure sustainable operation over extended periods, and it is crucial for the applications to be aware of their own energy consumption \cite{Hussain2017EnergyCO}.
Since many IoT devices are battery-powered or rely on energy harvesting, optimizing energy consumption is critical to prolonging their lifetime, reducing maintenance costs, and improving the overall network efficiency \cite{9370135}.
Power consumption in IoT devices is often determined by their operational profile, which includes multiple tasks such as sensing, communication, data processing, and idle periods. Different tasks may consume varying amounts of power, depending on their specific energy requirements and operational durations. 
Consequently, accurately modeling the power consumption for different tasks is essential for designing low-power IoT devices.

In \cite{Katz_2024}, the SUPERIOT\footnote{\url{https://superiot.eu/}} project, an initiative of the EU Smart Networks and Services Joint Undertaking (SNS JU), was introduced. The SUPERIOT project laid the foundation for developing sustainable IoT systems by proposing a comprehensive approach that integrates radio and optical technologies for dual-mode wireless connectivity, energy harvesting, and node positioning. The project also explored the use of sustainable materials such as printed electronics (PE) and conductive inks to minimize environmental impact across the lifecycle of IoT devices.
Building upon this foundational work, our current research dives deeper into the critical aspect of energy analysis, modeling, and prediction for the Silicon (Si)-based reconfigurable IoT (RIoT) node developed within the first phase of SUPERIOT framework. In the second phase, a hybrid node will be developed which will consist of a carrier board for the Si-based node developed in phase one, with flexible solar cell, energy harvesting manager integrated circuit (IC), supercapacitor for energy storage and accelerometer IC. In the third and final phase of the project, a fully printed node architecture will be implemented, such that it can be realized with only PE technology, namely printed organic photovoltaic  (OPV) cells and storage micro supercapacitor ($\mu$SC) designed using bio-based sustainable substrate (regenerated cellulose) components, similar to the work in \cite{Juha_paper}.

The focus of this study is on developing measurement-based models that accurately estimate energy consumption across various configurations and operating states of Silicon (Si)-based Reconfigurable IoT (RIoT) nodes, including Bluetooth Low Energy (BLE) communication, Narrow-Band Visible Light Communication (NBVLC), sensing, and E-ink displaying. Real-world energy consumption data has been utilized to validate these models, and multiple software and hardware optimizations have been incorporated to enhance the node's overall energy efficiency. 
This study not only aligns with the original goals of the SUPERIOT project but also adds tangible methods for energy optimization and predictive analysis, addressing the growing need for efficient energy management in the IoT ecosystem.

\section {Related Works}
While some studies have compared the energy consumption of widely used wireless technologies in IoT systems, including Wi-Fi, BLE, ZigBee, and UWB (ultra-wideband)  \cite{ comparison2, comparison3,7878777}, 
others have focused on developing models and strategies to reduce energy consumption in IoT networks.
For instance, the work in \cite{s18072104} specifically focuses on energy consumption modeling based on LoRa and LoRaWAN (Long Range Wide Area Network). It provides an approach to estimate the consumed power of each sensor node element within these communication technologies.
The authors of \cite{GARCIAMARTIN2023109855} developed a  power consumption model for estimating battery lifetime in narrowband IoT (NB-IoT) devices. Their model, based on the user equipment state diagram, incorporates mechanisms like extended discontinuous reception and power saving mode. 
The authors of \cite{OZKAYA2024105009} present a system-level energy consumption modeling approach for IoT devices. Their work enables fully simulated energy estimation of IoT applications without the need for complete design, implementation, or measurements. 
The authors of \cite{PowerofModels} developed a comprehensive model for estimating the power consumption of wireless sensor nodes in IoT applications. 
The proposed model takes a system-level perspective and accounts for all energy expenditures, including communications, data sensing, and processing. 
In \cite{4124316}, the authors propose an approach to model the energy consumption of embedded systems comprising a micro-controller, RAM memory, and an ADC. They analyze the software tasks executed by the micro-controller using a profiling procedure to gather information on system component activations and the instructions used. By employing an appropriate instrumentation setup, they model the energy behavior of each system component during both active and standby states. The total energy consumption is then determined by summing the energy contributions of the individual components. 
In \cite{7298286}, the authors address the challenge of designing energy-efficient IoT devices by focusing on three key building blocks: ultra-low power hardware platforms that integrate computing, sensing, storage, and wireless connectivity in a compact form factor; intelligent system-level power management techniques to optimize energy consumption; and the use of environmental energy harvesting to enable self-powered IoT devices.
\begin{figure}[t]
    \centering
    \subfigure[]{\includegraphics[width=0.45\linewidth]{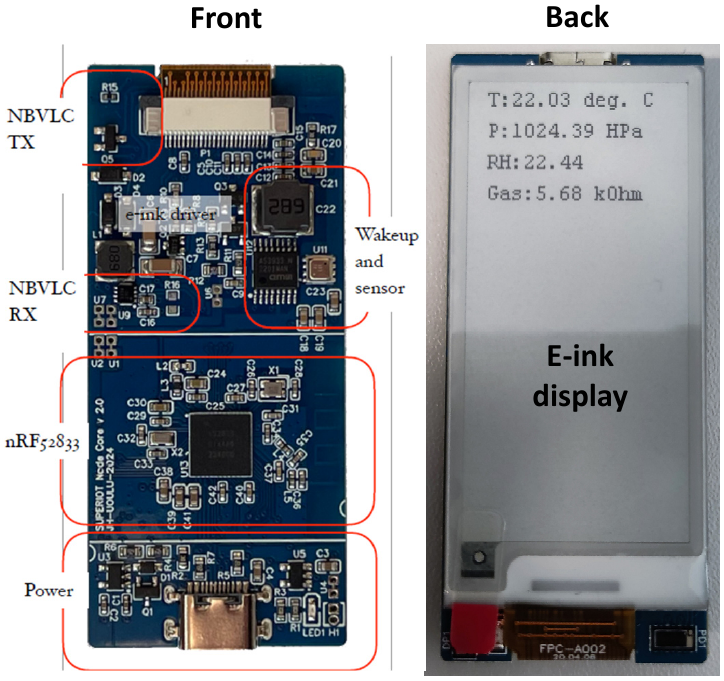}\label{fig:sibasednode}}
    \hspace{-0.7em}
    \subfigure[]{\includegraphics[width=0.40\linewidth]{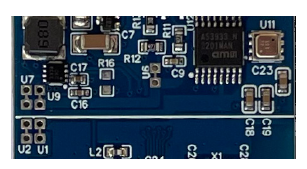}\label{fig:cutpoints}}
    \caption{Illustration of (a) Si-based RIoT node, (b) Cut points on node (see markings \textbf{U1}, \textbf{U2}, \textbf{U6}, \textbf{U7} and \textbf{U9}).}
    \label{fig:sinode}
    \vspace{-5mm}
\end{figure}

Despite these advances, there remains a need for measurement-based models that can accurately estimate the total energy consumption of IoT devices across various operational phases. Most existing models primarily focus on a specific task or scenario, such as processing, sensing or communication, without considering the entire sequence of operations a device undergoes during its lifetime. This paper addresses this gap by presenting  comprehensive models that estimate the energy consumption of the Si-based RIoT device developed in the first phase of the SUPERIOT project, including multiple operational phases: BLE communication, startup phase, regular cyclic operations involving sensing, E-ink display updates, VLC data transmission and reception, and idle periods. 
By analyzing energy consumption on the RIoT node, we are tailoring insights to the unique hardware configuration.
Our inclusion of E-ink display and NBVLC adds complexity, but it also opens up opportunities for novel energy-saving techniques.
We dive deeper into specific components (E-ink display, BLE, NBVLC) rather than treating the entire node as a black box.
Our  energy models can provide targeted optimizations for each feature.
By incorporating NBVLC, we explore a less common communication method, which could lead to energy savings compared to the traditional BLE.
The E-ink displays are known for their low power consumption because they only draw power when updating the screen content. Therefore, one of the many challenges lies in optimizing the update frequency and minimizing standby power.
The proposed models aim to provide an accurate representation of the device's power consumption over a given operating period, enabling better design and power management strategies. 
The proposed models provide insights into optimizing the power management strategies of IoT devices by quantifying the energy requirements of different operational modes.

\begin{figure}
    \centering    \includegraphics[width=1\linewidth]{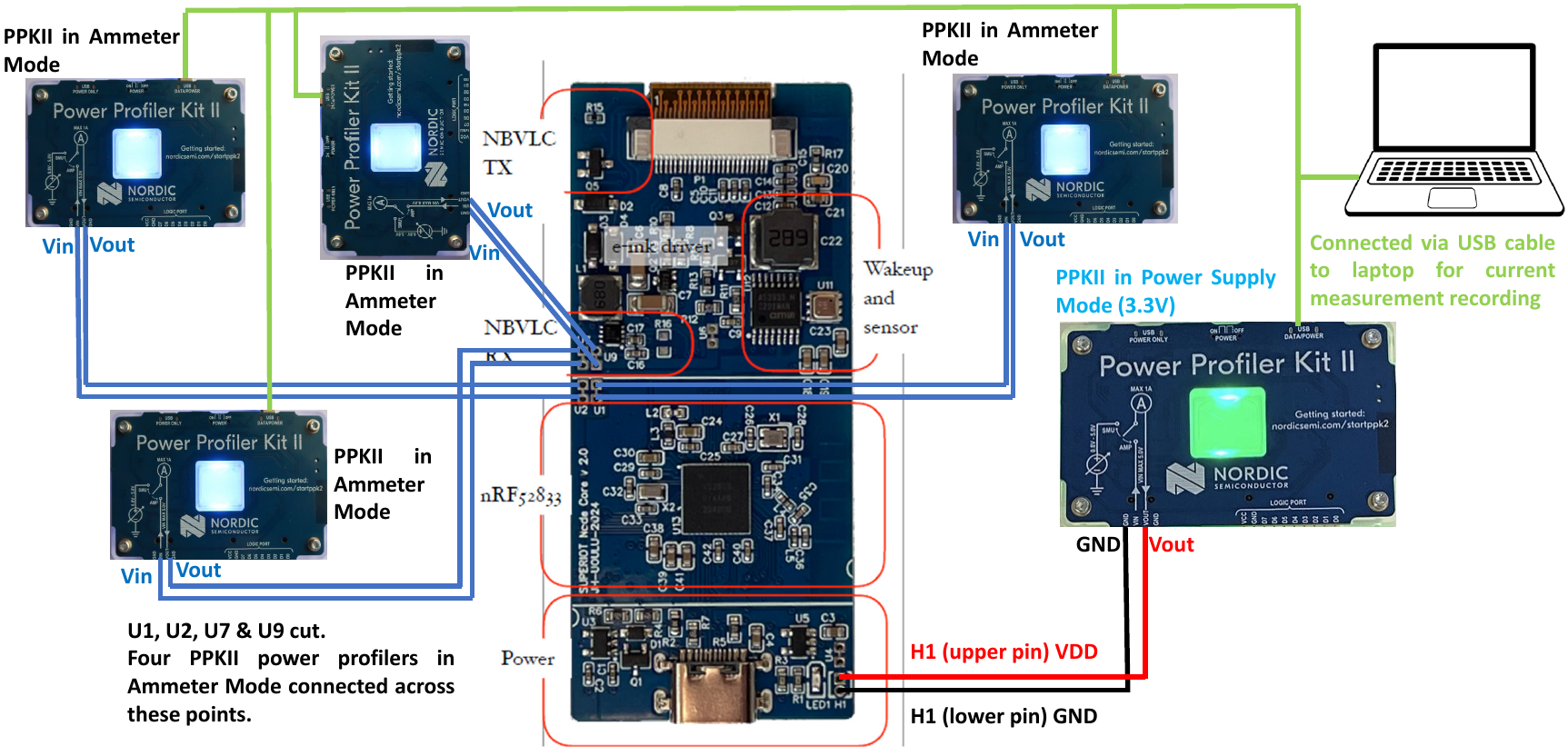}
    \caption{Current measurement setup. Points \textbf{U1}, \textbf{U2}, \textbf{U6}, \textbf{U7} and \textbf{U9} were cut, and jumper wires soldered across them to allow easy connection with power profilers. 
    }
    \label{fig:measurementsetup}
    \vspace{-5mm}
\end{figure}

\vspace{-3mm}
\section{Methodology}
\subsection{Overview of Si-based RIoT Node}
The Si-based based RIoT node developed in Phase 1 of the SUPERIOT project is shown in Fig. \ref{fig:sibasednode}. The node consists of the following key hardware components:
\begin{itemize}[leftmargin=*]
    \item \textbf{nRF52833 BLE SoC}: 
    This is a Bluetooth Low Energy (BLE) System on Chip (SoC) which serves as the main controller and communication interface for the node.
    
    \item \textbf{Narrow-band VLC RX/TX}: This component handles visible light communication (VLC), enabling data receiving and transmission through visible light (in this case infrared).
    The built-in NBVLC transmitter consists of infrared (IR) LED emitters (part no. XL-2012IRC-940) while the receiver is built upon the IR sensor preamplifier, VSOP38338, coupled with VEMD8080 photodiode.
    
    \item \textbf{BME688 sensor}: A versatile sensor capable of measuring environmental parameters such as temperature, humidity, pressure, and gas.

    \item \textbf{AS3933-BTST}: A low-power light-based wake-up and timer integrated circuit. This enhances power efficiency by enabling the system to wake up from a low-power state when needed based on light detection.
    
    \item \textbf{E-ink display}: 
    A 2.13-inch black/white E-ink display (250 $\times$ 122 pixels) with a full refresh rate of $\sim$2 s that provides a low-power method for visual output.
\end{itemize}

These components consume power during different operational phases. Each component has distinct power consumption characteristics, and they may not be active continuously. Instead, they operate in cycles, alternating between active operations and idle periods to conserve energy. For instance, a device might sense environmental data, display information on an E-ink screen, and communicate data using NBVLC and/or BLE, interspersed with idle periods. Understanding the energy consumption over these cycles is essential for designing energy-efficient IoT systems. Therefore, we perform direct measurement on the node to fully characterize the energy consumption of each component.

\vspace{-3mm}
\subsection{Energy Measurement Setup}
During the design phase, several cut (measuring) points were integrated on the Si-based RIoT node to facilitate current measurement across different parts of the node under normal operation. 
The cut points, illustrated in Fig. \ref{fig:cutpoints}, are shorted by default and need to be cut to enable current measurement across the core components. Specifically, cut point \textbf{U1} allows current measurement from the 3.3 V supply to the nRF52833, wake-up IC, and sensor. \textbf{U2} facilitates current measurement from the 3.3 V supply to the E-ink driver and the narrow-band VLC transmitter and receiver. \textbf{U7} is dedicated to measuring current to the NBVLC TX module only, while \textbf{U9} is for the NBVLC RX module only. Additionally, \textbf{U6} is not used for current measurement but can be cut to reduce sleep current when the E-ink display is powered off.

\begin{table}[t]
\centering
\caption{Current draw during an operation cycle consisting of sensing, E-ink displaying, and idle states while node is connected via BLE to a central device.}
\begin{tblr}{
  width = \linewidth,
  colspec = {Q[65]Q[103]Q[404]Q[178]Q[168]},
  cell{1}{2} = {c=4}{0.873\linewidth,c},
  hlines,
  vlines,
}
 & \textbf{Average current consumption (mA)} &  &  & \\
 \textbf{Point} & \textbf{Overall} & \textbf{Idle after E-ink display / Idle after sensing \quad($\sim$7.5 s) / ($\sim$200 ms)} & \textbf{Sensing ($\sim$516 ms) } & \textbf{E-ink display ($\sim$2.8 s)}\\
\textbf{U1} & 5.79 & 5.47   / 5.48 & 12.05 & 5.52\\
\textbf{U2} & 0.73 & 0.41   / 0.41 & 0.41 & 1.68\\
\textbf{U7} & 0.0003 & 0.0003 & 0.0003 & 0.0003\\
\textbf{U9} & 0.33 & 0.33  / 0.33 & 0.33 & 0.33
\end{tblr}
\label{Tab:sensing_eink_ble}
\vspace{-5mm}
\end{table}

To perform an in-depth energy analysis on the node, we utilized the Power Profiler Kit II (PPKII) by Nordic Semiconductor. 
Fig. \ref{fig:measurementsetup} shows the overall diagram of the hardware setup utilized in our current consumption analysis. 
Overall, five PPKII boards were used. One is used in power supply mode (3.3 V) to power the node through the \textbf{H1} pins and at the same time obtain current measurement across the whole node. The other four PPKII boards are set to ammeter mode and connected across the four cut points, namely, \textbf{U1}, \textbf{U2}, \textbf{U7} and \textbf{U9}. All PPKII boards were connected via USB to a laptop to record the current measurements using the \textit{nRF Connect for Desktop} application, available from Nordic Semiconductor.

\vspace{-3mm}
\subsection{Energy Measurement and Analysis}
After the Si-based RIoT node was designed and implemented, a firmware with different test modes was built for the node for testing the different core components and conducting power measurements on various sub-circuits. 
\begin{figure}[t]
    \centering
    \includegraphics[width=0.65\linewidth]{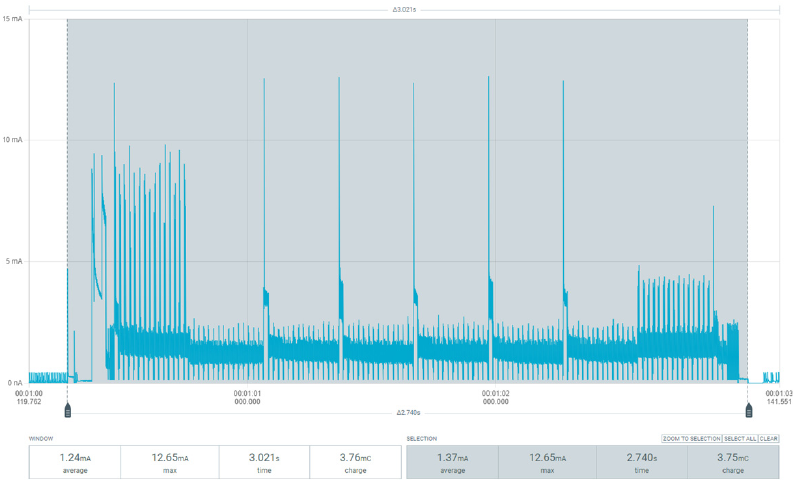}
    \caption{Current profile as measured across measuring point \textbf{U2} when unoptimized E-ink display was active.}
    \label{fig:eink}
    \vspace{-5mm}
\end{figure}

\subsubsection{Sensing and E-ink displaying test (while BLE connection active)}
In this experiment, the node is connected via BLE to a smartphone (central device) and the test mode is activated using the \textit{nRF Connect} app.
In this test, the node performs sensing using the integrated BME688 sensor for about 516 ms, displays the sensor readings on the E-ink display for $\sim$2.8 s and is idle for about 7.5 s (one cycle $\sim$11 s). Note by ``idle", we mean that the node is not doing any operations but remains connected to a central device via BLE with a default connection interval of 45 ms and BLE TX power level of 0 dBm. Table \ref{Tab:sensing_eink_ble} shows the average current consumption during these operations, as measured across the 
measuring points \textbf{U1}, \textbf{U2}, \textbf{U7} and \textbf{U9} (see Fig. \ref{fig:measurementsetup}). In this test mode, the sensor and E-ink display are active in branches \textbf{U1} and \textbf{U2}, respectively. The NBVLC TX and NBVLC RX are not used (inactive) in this firmware test mode and hence in Table \ref{Tab:sensing_eink_ble} it can be observed that their current consumption remains essentially unchanged, at around 0.0003 mA and 0.33 mA, respectively. 
Considering the measuring point \textbf{U2} current profile in Table \ref{Tab:sensing_eink_ble}, 
it can be observed that the average current draw for the E-ink display when active is $\sim$1.37 mA (after subtracting the inactive current draw from the NBVLC TX and RX). This value aligns with the one measured in Fig. \ref{fig:eink}, where an average current of $\sim$1.37 mA is recorded over a period of $\sim$2.8 s across measuring point \textbf{U2} when the E-ink display was active. Note that to get the current across the E-ink display only, points \textbf{U7} and \textbf{U9} were cut during the test. When the E-ink display is inactive, the average current flowing through it is around $\sim$83 uA.

\subsubsection{NBVLC TX and NBVLC RX test (while BLE connection active)}
In this test, the node performs NBVLC transmission (TX) and reception (RX) every 5 s. During the idle period (i.e. no NBVLC TX and RX), the node remains connected to the central device via BLE with a connection interval of 45 ms and BLE TX power of 0 dBm. Note that in this firmware test mode, a NBVLC message is transmitted using the built-in IR-LED transmitter and the node receives that same message with the built-in NBVLC receiver. The E-ink display and sensor were inactive in this experiment. Table \ref{tab:nbvlc} shows the average current consumption for this firmware test mode during the active state (i.e. during NBVLC transmission and reception) and idle state. 
The change in current during NBVLC transmitting and receiving is more evident in current profiles of \textbf{U2}, \textbf{U7} and \textbf{U9}. The peak rise in current for the NBVLC transmission and reception happens for around  $\sim$84 ms. During this time, the NBVLC TX consumes a current of  $\sim$2.6 mA compared to its idle current of 0.35 uA. For the NBVLC RX, the current consumption is less substantial ($\sim$430 uA when active compared to its idle current of $\sim$330 uA).

Based on these two test modes, it is clear that BLE is the most power-hungry operation in the node, accounting for over 90\% of the overall current consumption. 
This behavior is evident by observing the average idle current across \textbf{U1}, and comparing it with the average current across the other cut points, namely, \textbf{U2}, \textbf{U7}, and \textbf{U9}.
In contrast, the E-ink display, NBVLC TX, and NBVLC RX components consume minimal current during their inactive states, with the NBVLC TX showing the lowest consumption when idle. However, when the NBVLC module is activated, there is a sharp spike in current consumption, particularly during NBVLC transmission, though this surge is brief. The NBVLC receiver, while also showing an increase in current draw when active, exhibits a more moderate rise of approximately 30\%. Among these three components, the E-ink driver has the highest energy consumption during operation, largely due to its prolonged refresh or active period compared to the brief bursts of the NBVLC TX and RX.

\vspace{-3mm}
\section{Si-based Node optimization}
\subsection{Software optimization}
\label{software_opt}
In this section, we perform software-level optimization in order to reduce the overall current consumption of the node while it is performing several of its core functions, namely, BLE, sensing, and E-ink displaying. However, in this case, we found out that certain timer functions in the firmware code, which are typically used for NBVLC, significantly increase the overall current consumption when in use. 

\begin{table}
\centering
\centering
\captionsetup{}
\caption{Current draw during during an operation cycle consisting of NBVLC TX/RX and idle state while node is connected via BLE to a central device.}
\begin{tblr}{
  width = \linewidth,
  colspec = {Q[83]Q[115]Q[296]Q[444]},
  cell{1}{2} = {c=3}{0.855\linewidth,c},
  hlines,
  vlines,
}
 & \textbf{Average current comsumption (mA)} &  & \\
 \textbf{Point} & \textbf{Overall~} & \textbf{Idle~($\sim$5.31 s)} & \textbf{~NBVLC TX/RX~($\sim$84 ms)}\\
\textbf{U1} & 5.42 & 5.42 & 5.42\\
\textbf{U2} & 0.46 & 0.41 & 3.8\\
\textbf{U7} & 0.040 & 0.00035 & 2.56\\
\textbf{U9} & 0.33 & 0.33 & 0.43
\end{tblr}
\label{tab:nbvlc}
\vspace{-5mm}
\end{table}
\begin{figure}[t]
    \centering
    \includegraphics[width=0.65\linewidth]{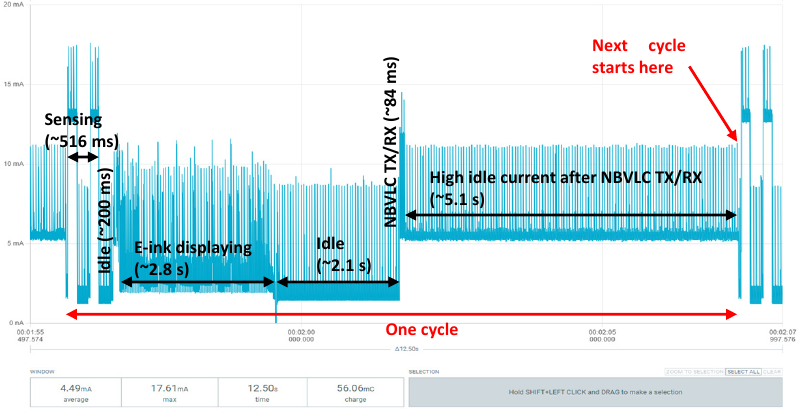}
    \caption{Current profile as measured across node during low power sensing, E-ink displaying (unoptimized) and NBVLC TX/RX while node is connected to a central device via BLE.  \textit{\textbf{nbvlc\_off()}} function is not activated after NBVLC data reception.}
    \label{fig:lowpower_sen_eink_nbvlc}
    \vspace{-5mm}
\end{figure}
The NBVLC functionality is implemented using timers and interrupts. A function (defined as \textbf{\textit{TimerHandler0}}) in the firmware  code handles the timing and modulation required for IR communication. The code follows the NEC protocol for encoding and decoding infrared signals. This includes start bits, data bits, and error checking. The NEC frame consists of a 32-bit packet with address and data fields. 
Another function, namely a reception handler monitors incoming signals, checking for correct intervals and interpreting them as data. The code uses an NRF52 timer to schedule operations, with a given interval (as defined by a constant value \textbf{\textit{TIMER0\_INTERVAL\_MS}}. A nRF52 PWM instance is also defined in the code to control the duty cycle of the Pulse Width Modulation (PWM) signal sent to the IR (NBVLC) transmitter. Two functions, \textbf{\textit{nbvlc\_on()}} and \textbf{\textit{nbvlc\_off()}}, control the timer responsible for handling NBVLC, and basically switch the NBVLC functionality on or off.
The primary purpose of \textit{\textbf{nbvlc\_on()}} in Listing~\ref{lst:nbvlc_functions} is to enable the timer to call \textbf{\textit{TimerHandler0}} every 50 ms (\textit{\textbf{TIMER0\_INTERVAL\_MS}} is defined as 500 ms). It calls the \textit{\textbf{TimerHandler0}} at a higher frequency, enabling fast PWM control of the transmission process for sending IR data.
The \textit{\textbf{nbvlc\_off()}} function shown in Listing~\ref{lst:nbvlc_functions} modifies the interval at which \textit{\textbf{TimerHandler0}} is called. By multiplying the original interval (500 ms) by 10, the timer's interval is significantly increased, making the \textit{\textbf{TimerHandler0}} function execute less frequently. By increasing the interval of the timer in \textit{\textbf{nbvlc\_off()}} and reducing the frequency at which \textit{\textbf{TimerHandler0}} is called, we are effectively decreasing the workload and the frequency of operations in the system. This reduction in activity can lead to lower power consumption because the processor and other components have less work to do and can spend more time in low-power states.
In order to obtain low-power sensing and E-ink displaying while the node is connected via BLE to a central device, the \textit{\textbf{nbvlc\_off()}} function is included in the firmware code just before the sensing and E-ink display operations are activated.

\begin{lstlisting}[style=arduinoStyle, caption={\textit{\textbf{nbvlc\_on()}} and \textit{\textbf{nbvlc\_off()}} functions.}, label={lst:nbvlc_functions}]
void nbvlc_on() 
{
  ITimer0.setInterval(TIMER0_INTERVAL_MS / 10, TimerHandler0);
}
void nbvlc_off()
{
  ITimer0.setInterval(TIMER0_INTERVAL_MS * 10, TimerHandler0);
}
\end{lstlisting}

When the NBVLC functionality is activated in addition to BLE, sensing, and E-ink operations, the \textit{\textbf{nbvlc\_on()}} function must be invoked before the node initiates NBVLC data transmission. 
However, 
the idle current following NBVLC TX/RX will remain elevated, as illustrated in Fig. \ref{fig:lowpower_sen_eink_nbvlc}.
To mitigate this, it is essential to call the \textit{\textbf{nbvlc\_off()}} function immediately after 
the NBVLC data reception is executed.
Then, in this case we will obtain a low-power consumption across all operation states as shown in Fig. \ref{fig:lowpower_unopteink}.
Note that the idle period in between the different operations such as sensing, E-ink displaying, and NBVLC TX/RX can be modified in the firmware code as per user's requirements. 
It is also noteworthy that the NBVLC TX/RX in 
Fig. \ref{fig:lowpower_unopteink}
occurs over an average duration of 100 ms (average current consumption is approximately 8.2 mA), as opposed to $\sim$84 ms (see Fig. \ref{fig:unopt}). This increase is anticipated, as the node requires additional time to transition into low-power mode following the completion of NBVLC data reception.

Note that the \textit{\textbf{nbvlc\_off()}} function can also be added in the BLE \textit{\textbf{connect\_callback()}} function in the firmware Arduino code to obtain a low power consumption when the node is in BLE connected-only mode with
a central device. 
In a BLE application, this function is invoked when a connection event occurs between the node (in our case the node acts as a peripheral device) and a central device. Inside this function, one can typically handle the tasks that need to occur right after the BLE connection is established. 
Consequently, when a central device connects to the node, the latter transitions to a significantly lower current consumption, 
achieving a 74\% decrease in average current consumption, considering a BLE connection interval of 45 ms and TX power level of 0 dBm (see Table \ref{tab:ble_conn_txpower}).

\begin{figure}[t]
    \centering    \includegraphics[width=0.75\linewidth]{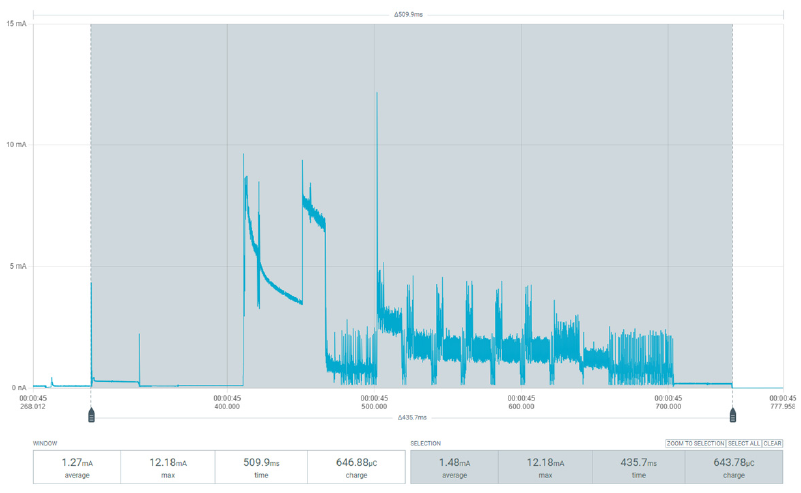}
    \caption{Current profile as measured across measuring point \textbf{U2} when optimized E-ink display was active.}
    \label{fig:einkopt}
    \vspace{-5mm}
\end{figure}

\vspace{-3mm}
\subsection{E-ink display optimization}
In order to further reduce the energy consumption of the node during its operation, the E-ink display was also optimized. Typically, the Look Up Table (LUT) in the E-ink display firmware was refined to achieve low power when it is active. The waveform is a LUT for the E-ink controller, guiding how it drives the display pixels \cite{5418847}. 
The initial (unoptimized) configuration of the E-ink display exhibits an average current consumption of approximately 1.4 mA over a duration of around 2.8 seconds at measurement point \textbf{U2}, as shown in Fig. \ref{fig:eink}. This occurs when the E-ink display is active.
In the low-power optimized configuration, shown in Fig. \ref{fig:einkopt}, the E-ink display records an average current of approximately 1.5 mA over a shorter active period of about 435 ms at the same measurement point \textbf{U2}.
Thus, the refresh time of the E-ink display is reduced by 84\% while the energy consumption is reduced from 12.39 mJ to 2.13 mJ (83\% reduction).

\begin{figure*}[t]
    \centering
    \subfigure[]{\includegraphics[width=0.32\linewidth]{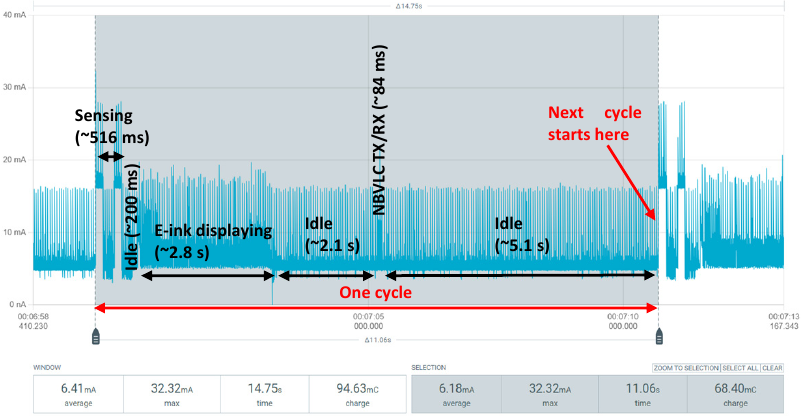}\label{fig:unopt}}
    \hspace{0pt}
    \subfigure[]{\includegraphics[width=0.32\linewidth]{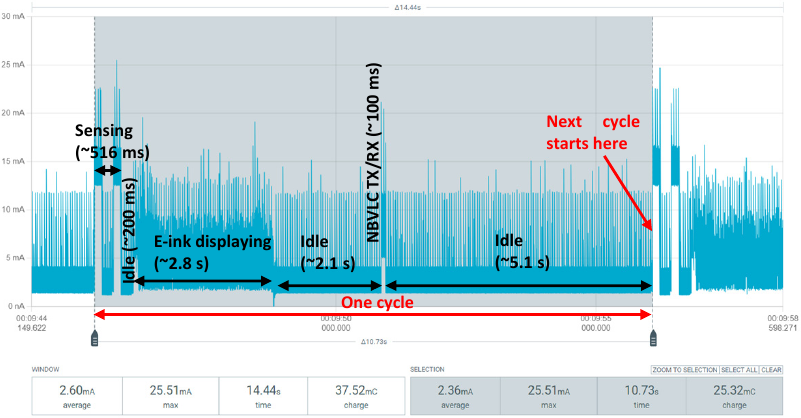}\label{fig:lowpower_unopteink}}
    \hspace{0pt}
    \subfigure[]{\includegraphics[width=0.32\linewidth]{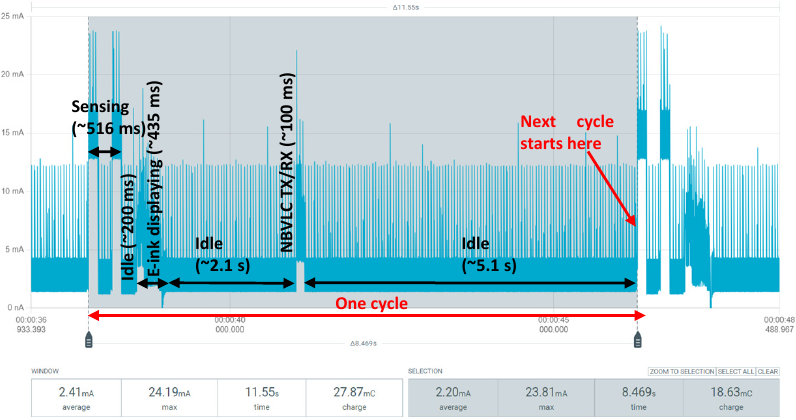}\label{fig:lowpower_opteink}} 
    \caption{Average current consumption measured across node when all functionalities are enabled: (a) Normal (unoptimized) node operation with unoptimized E-ink display, (b) Low-power (software optimized) node operation with unoptimized E-ink display, and (c) Low-power (software optimized) node operation with optimized E-ink display.}
    \label{fig:lowpowerwitheinkopt}
    \vspace{-5mm}
\end{figure*}

With the optimized E-ink display, the energy consumption of the node can be further reduced, in addition to the software optimizations performed in Section \ref{software_opt}.
Let's consider the node’s average current consumption when all functionalities—BLE, sensing, E-ink display, and NBVLC TX/RX—are enabled during an operation cycle. Fig. \ref{fig:lowpowerwitheinkopt} presents the average current consumption for three test scenarios.
\begin{itemize}[leftmargin=*]
    \item Fig. \ref{fig:unopt} shows the current profile for the node operating in a normal, unoptimized configuration. In this scenario the node performs sensing, E-ink displaying, and NBVLC TX/RX in a single operation cycle while maintaining a BLE connection with a central device.    
    \item Fig. \ref{fig:lowpower_unopteink} illustrates the current profile for the node operating with low-power, software-level optimizations (see Section \ref{software_opt}) while performing the same operations. In both Fig. \ref{fig:unopt} and Fig. \ref{fig:lowpower_unopteink}, the initial, unoptimized E-ink display is used.    
    \item Fig. \ref{fig:lowpower_opteink} depicts the same scenario as in Fig. \ref{fig:lowpower_unopteink}, but with the optimized E-ink display in use.
\end{itemize}
In all three cases, an identical idle period is configured after sensing, E-ink displaying, and NBVLC TX/RX. The results show that software-level optimizations reduce the average current consumption significantly during an operation cycle—from 6.18 mA in the unoptimized state (Fig. \ref{fig:unopt}) to 2.36 mA (Fig. \ref{fig:lowpower_unopteink}). Further reductions are achieved by optimizing the E-ink display, lowering the average current to 2.20 mA (Fig. \ref{fig:lowpower_opteink}).
Given an operating voltage of 3.3 V, the energy consumption per cycle for the unoptimized node is approximately 0.2257 J. For the low-power, software-optimized node with the optimized E-ink display, the energy consumption drops to about 0.0615 J. This corresponds to a 73\% reduction in energy consumption during an operation cycle involving sensing, E-ink displaying, and NBVLC TX/RX while the node is connected via BLE (i.e. all functionalities enabled).

\vspace{-3mm}
\subsection{Hardware optimization}
\label{hardware_opt}
In this section, a very low power firmware is used, where the node periodically updates sensor readings to the E-ink display, with the updates occurring approximately every two minutes. However, in this firmware neither BLE nor VLC communication are active. After each update, the core enters a deep sleep state.
Approximately two minutes later, the AS3933 wake-up IC generates a pulse, causing the nRF52833 SoC to wake up and reboot. As a result, no data is retained during the sleep period. Due to the circuitry setup, power consumption during sleep is attributed to the nRF52833 SoC, the AS3933 wake-up IC, and the BME688 environmental sensor. To further reduce current consumption during deep sleep, the \textbf{U9} and \textbf{U6} points (see Fig. \ref{fig:cutpoints}) should be cut.
\textbf{U6} is not used for current measurement but can be cut to reduce the deep sleep current while E-ink display is powered off. Thus, we analyzed the current consumption of the very low power firmware under the following four hardware optimization configurations:
\begin{enumerate}[]
    \item points \textbf{U6} \& \textbf{U9} shorted (default configuration).
    \item point \textbf{U6} cut only.
    \item point \textbf{U9} cut only.
    \item both points \textbf{U6} and \textbf{U9} cut.
\end{enumerate}

\begin{figure}[t]
    \centering
    \subfigure[]{\includegraphics[width=0.65\linewidth]{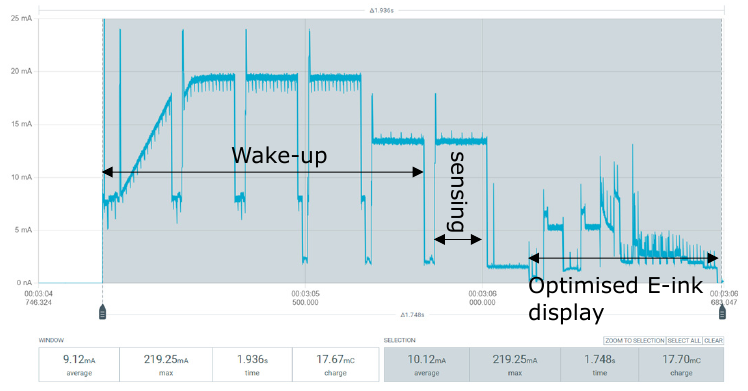}\label{fig:active}}
    \hfill
    \subfigure[]{\includegraphics[width=0.65\linewidth]{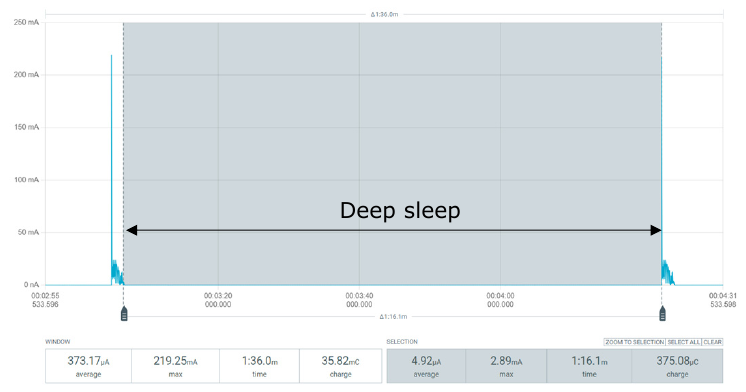}\label{fig:deepsleep}}
    \caption{Current profile of the very low power firmware as measured across the node during: (a) active phase, and (b) deep sleep phase (\textbf{U6} and \textbf{U9} both cut).}
    \label{fig:verylowpower}
    \vspace{-3mm}
\end{figure}

\begin{figure}[t]
    \centering
    \subfigure[]{\includegraphics[width=0.5\linewidth]{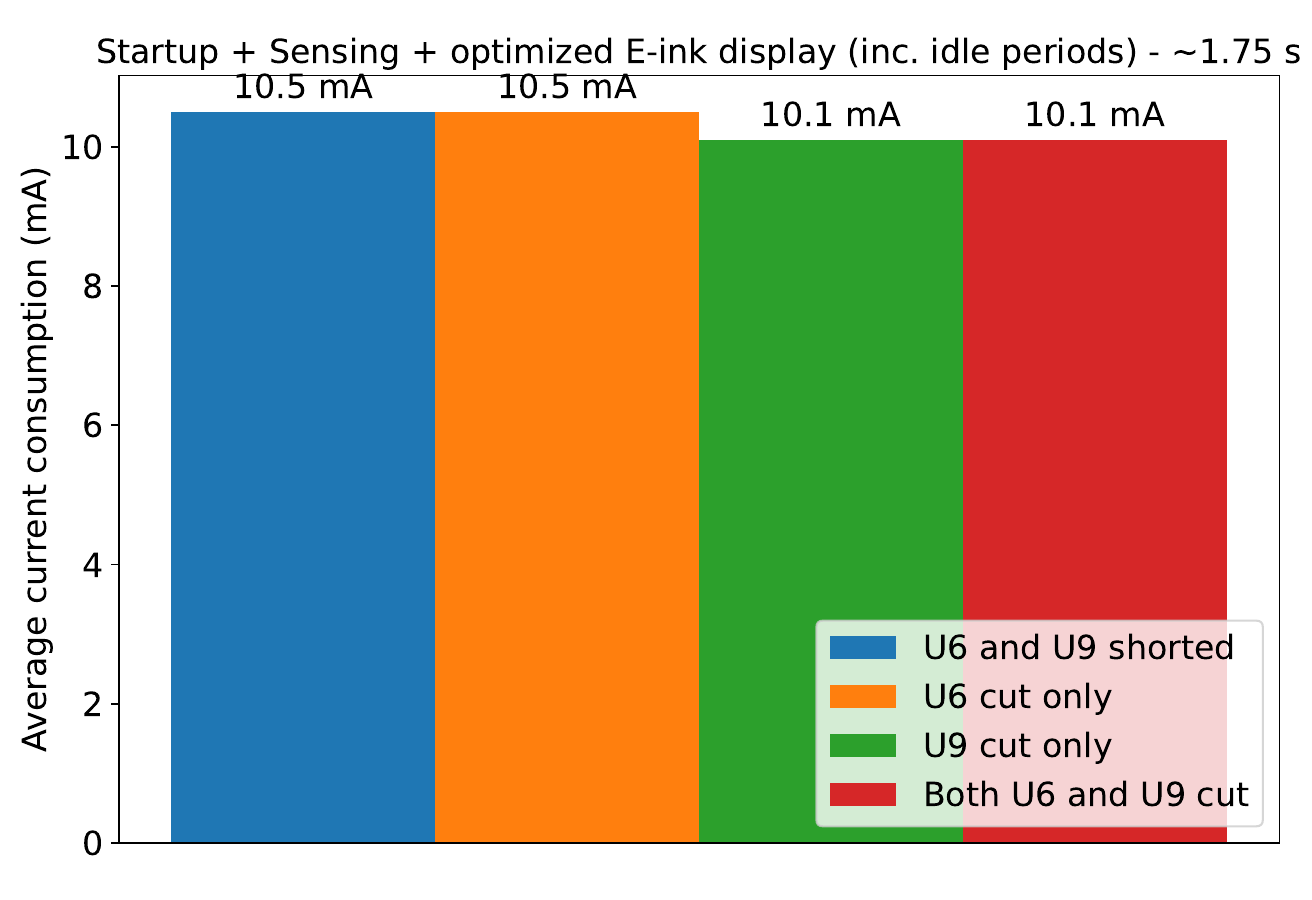}\label{fig:active2}}
    \hspace{-0.9em}  
    \subfigure[]{\includegraphics[width=0.5\linewidth]{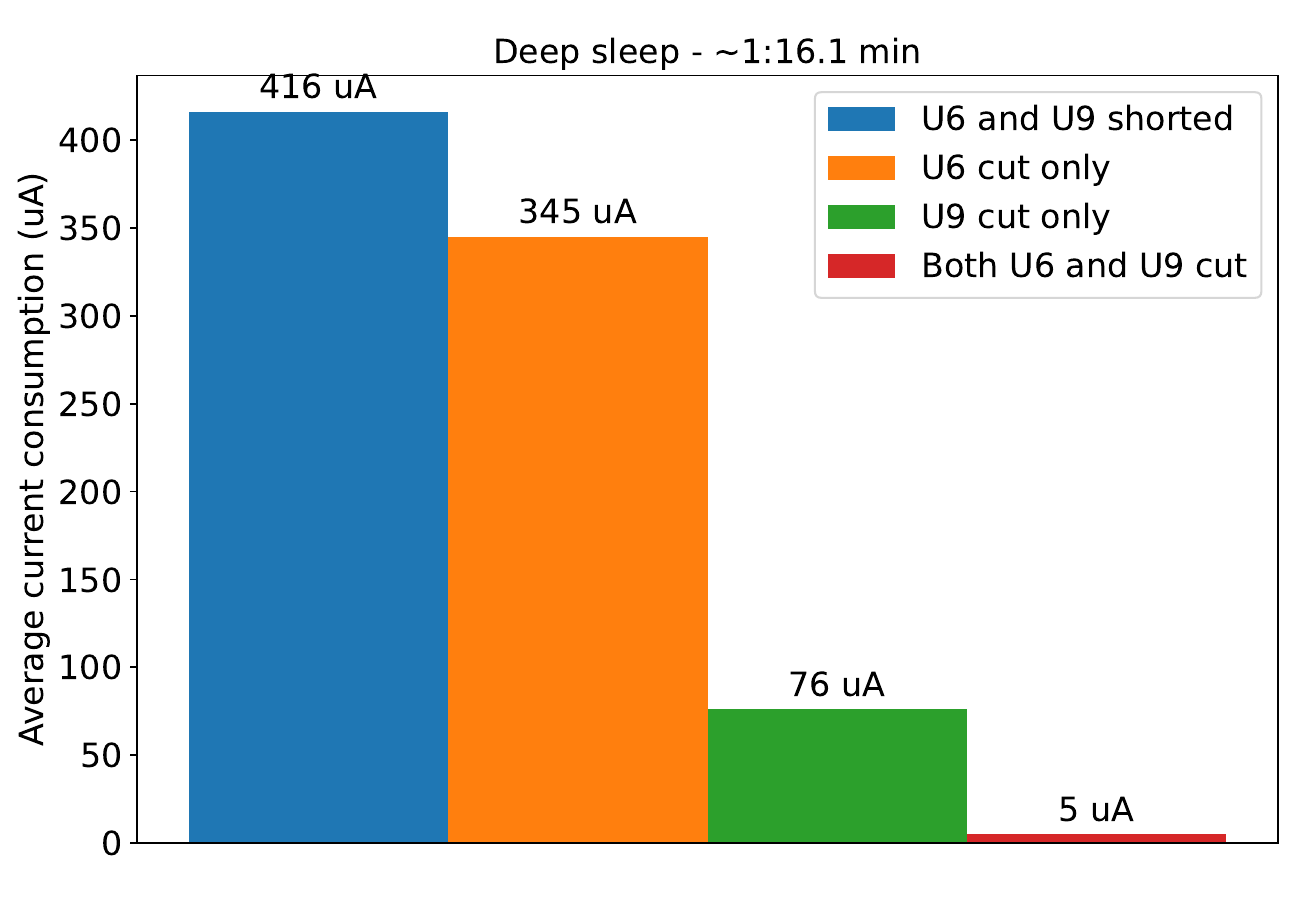}\label{fig:deepsleep2}}
    \caption{Average current consumption of the very low power firmware (with BLE and VLC both turned off) measured across the node during: (a) active phase, (b) deep sleep phase.}
    \label{fig:verylowpower_active_deepsleep}
    \vspace{-5mm}    
\end{figure}

\begin{figure}[t]
    \centering
    \includegraphics[width=0.6\linewidth]{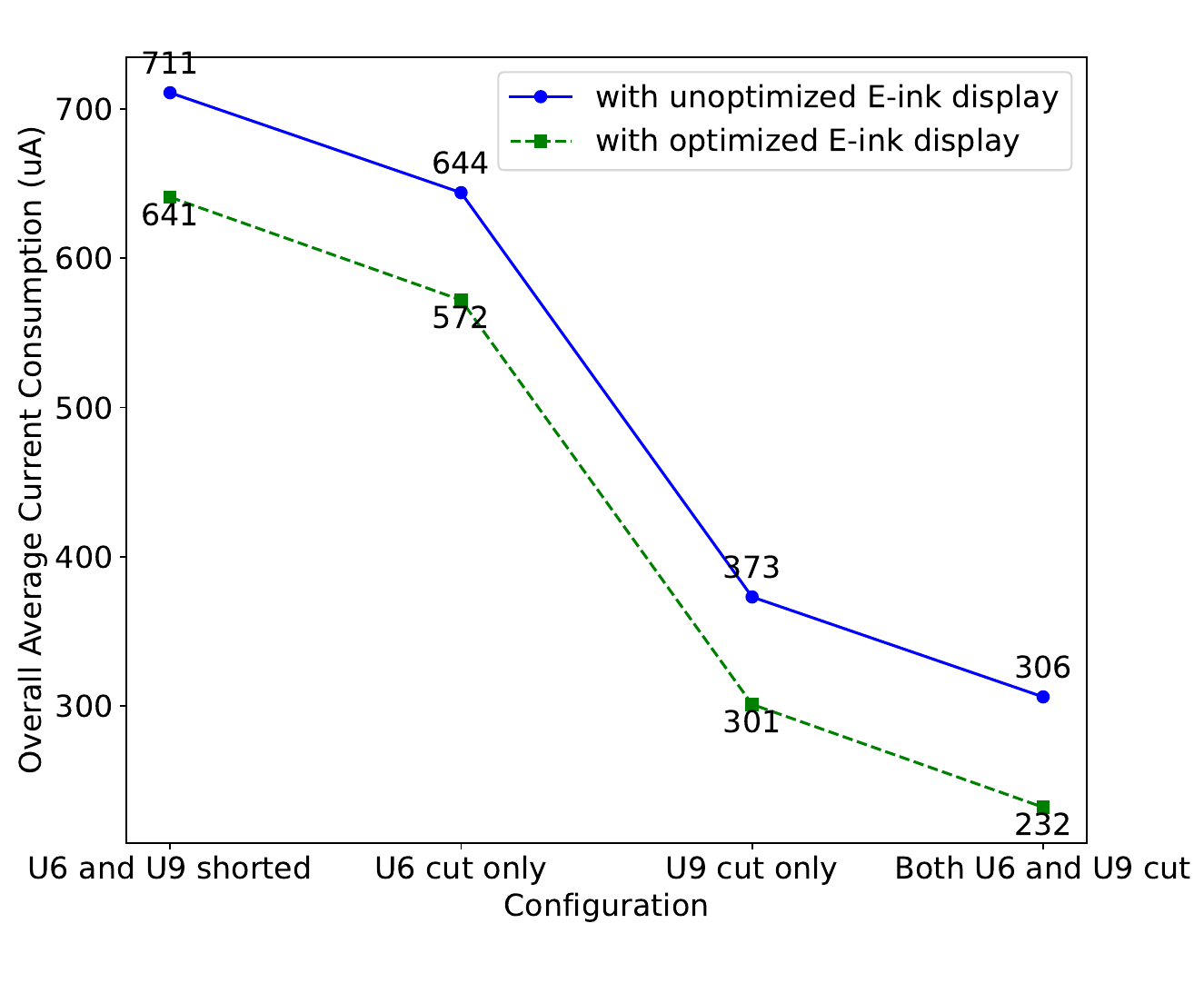}
    \caption{Overall average current consumption of the very low power firmware (with BLE and VLC both turned off) measured across the node considering both the active and deep sleep phases.}
    \label{fig:verylowpower_overall}
    \vspace{-5mm}
\end{figure}

Fig. \ref{fig:verylowpower} shows the current profile of the very low power firmware when measured across the node during the active phase (Fig. \ref{fig:active}) and deep sleep phase (Fig. \ref{fig:deepsleep}). In this case, both points \textbf{U6} and \textbf{U9} were cut. Sensing and optimized E-ink displaying are performed immediately as soon as the SoC wakes up from deep sleep. Table \ref{tab:verylowpower} shows the average current measured across the whole node for each operation separately, when the node is in very low power mode.   
Fig. \ref{fig:verylowpower_active_deepsleep} shows the average current consumption of the very low power firmware as measured across the 
node during active (startup, sensing, E-ink displaying, including any idle period between them) phase and deep sleep phases.
The startup, sensing and E-ink display operations (active state) take around 1.8 s to execute before the node goes in a deep sleep state (for around 1:16.1 minutes).
Considering Fig. \ref{fig:active2}, the average current remains consistent at 10.5 mA when both \textbf{U6} and \textbf{U9} are shorted or when only \textbf{U6} is cut. There is a slight decrease (by approx. 4\%) to 10.1 mA when only \textbf{U9} is cut or when both \textbf{U6} and \textbf{U9} are cut. This means that the impact of cutting \textbf{U6} in this case is negligible.
As for the deep sleep phase in Fig. \ref{fig:deepsleep2}, we observe that there is a substantial reduction in deep sleep current consumption (by approx. 99\%) when cutting either \textbf{U6}, or \textbf{U9}, or both (from 416 uA to 5 uA). 
Cutting \textbf{U6} only has no impact on the current consumption across the node during the active phase. However, for the deep-sleep phase, cutting \textbf{U6} results in a lower current across the node.
Thus, we can conclude that cutting \textbf{U6} and \textbf{U9} have clear benefits in terms of current consumption and will lead to energy optimization (as can be observed in Fig. \ref{fig:verylowpower_overall}). These two cuts (\textbf{U9} and \textbf{U6}) minimize the sleep currents of the NBVLC receiver and E-ink display, respectively.
In case of the NBVLC RX (\textbf{U9}), the cut isolates the receiver circuit from common VDD. In this case, one must solder in a 0 Ohm resistor to allow powering of the circuitry by switching the relevant nRF52833 GPIO pin to HIGH during normal operation, and to shut down the circuit by switching the pin to LOW during sleep.
From Fig. \ref{fig:verylowpower_overall}, we can also deduce that the reduction in the E-ink display refresh (active) time further decreases the overall average current consumption across the node, under various hardware optimization configurations.

\begin{table}[t]
\centering
\caption{Average current measured for each operation in the very low-power firmware with both VLC and BLE disabled.}
\begin{tblr}{
  width = \linewidth,
  colspec = {Q[408]Q[158]Q[104]Q[104]Q[163]},
  row{1} = {c},
  cell{1}{1} = {r=2}{},
  cell{1}{2} = {c=4}{0.529\linewidth},
  vlines,
  hline{1,3-9} = {-}{},
  hline{2} = {2-5}{},
}
\textbf{Node operations (duration)} & \textbf{Average current across node (mA)} &  &  & \\
 & \textbf{U6 and U9 shorted} & \textbf{U6 cut only} & \textbf{U9 cut only} & \textbf{Both U6 and U9 cut}\\
Startup
  ($\sim$909 ms) & 15.5
  & 15.5
  & 15.1 
  & 
  15.1 
  \\
Idle
  between startup and sensing ($\sim$30 ms) & 2.4
  & 2.4 
  & 2.1
  & 2.1
  \\
Sensing
  ($\sim$149 ms) & 13.8
  & 13.8 
  & 13.4 
  & 13.4 
  \\
Idle
  between sensing and E-ink display ($\sim$118 ms) & 1.9
  & 1.9 
  & 1.6
  & 1.6
  \\
optimized E-ink ($\sim$544 ms) / Unoptimized E-ink ($\sim$2.8 s) & 3.53 / 2.9
  & 3.53 / 2.9   
  & 3.2 / 2.6
  & 3.2 / 2.6
  \\
Deep
  sleep after E-ink display ($\sim$1:16.1 min) & 0.416 
  & 0.345
  & 0.076
  & 0.005
\end{tblr}
\label{tab:verylowpower}
\vspace{-5mm}
\end{table}

\begin{figure}[t]
    \centering
    \includegraphics[width=0.75\linewidth]{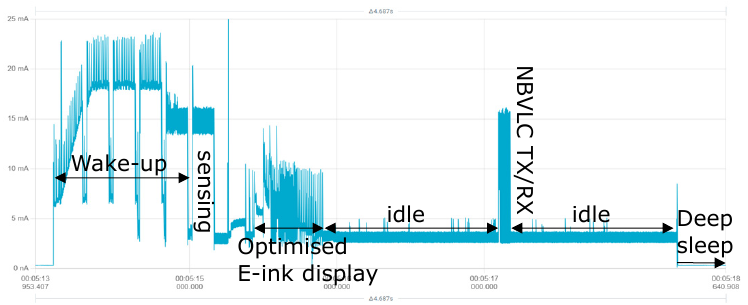}
    \caption{Current profile of the very low power firmware with NBVLC TX/RX functionality enabled (BLE off), as measured across the node (\textbf{U6} cut, \textbf{U9} shorted).}
    \label{fig:verylowpower_nbvlc}
    \vspace{-3mm}
\end{figure}

Furthermore, we examined the average current consumption of the very low-power firmware with the NBVLC functionality enabled, alongside other operations such as startup, sensing and E-ink display. It is important to note that the BLE functionality remained disabled during this test. The current consumption was analyzed under two specific conditions with the NBVLC functionality active:
\begin{enumerate}[]
    \item \textbf{U6} and \textbf{U9} points shorted.
    \item \textbf{U6} point cut only.
\end{enumerate}
In this test, we did not cut the \textbf{U9} measuring point, as doing so would disable the NBVLC receiver (RX) functionality. 
Fig. \ref{fig:verylowpower_nbvlc} shows the current profile of the low power firmware with NBVLC functionality enabled. Note that in this case, we added a delay (idle period) of 1 s in the firmware after the E-ink display
and
NBVLC TX/RX. 
Table \ref{tab:verylowpower_nbvlc} shows the average current measured across the node for each operation when the node is in very low power mode (with NBVLC functionality enabled). 
Comparing Table \ref{tab:verylowpower_nbvlc} with Table \ref{tab:verylowpower}, 
we observe that the activation of the NBVLC functionality has increased the average current consumption across the operations in common (excluding the deep sleep current which remains essentially unchanged).
With the NBVLC module active, the BLE SoC (nRF52833) increases its overall workload (higher CPU usage) as it might require more coordination between the different subsystems, increasing the complexity of operations and, thus, energy consumption. 
Additionally, we note that the idle periods before and after the NBVLC TX/RX (see Fig.  \ref{fig:verylowpower_nbvlc} and Table \ref{tab:verylowpower_nbvlc}) exhibit high current consumption. As a result, these idle periods should be carefully defined. 
In Table \ref{tab:verylowpower_nbvlc}, we also note that cutting \textbf{U6} lowers the overall current consumption during the operation cycle.

\begin{table}[t]
\centering
\caption{Average current measured across node for each operation with NBVLC functionality enabled in the very low power firmware (BLE off).}
\begin{tblr}{
  width = \linewidth,
  colspec = {Q[585]Q[212]Q[162]},
  row{1} = {c},
  cell{1}{1} = {r=2}{},
  cell{1}{2} = {c=2}{0.414\linewidth},
  vlines,
  hline{1,3-13} = {-}{},
  hline{2} = {2-3}{},
}
\textbf{Node operations} & \textbf{Average current across node (mA)} & \\
 & \textbf{U6 and U9 shorted} & \textbf{U6 cut only}\\
Startup
  ($\sim$910 ms) & 15.8
   & 14.6
  \\
Idle
  between startup and sensing ($\sim$30 ms) & 4.84
   & 3.53
  \\
Sensing
  ($\sim$149 ms) & 16  & 14.8 \\
Idle
  between sensing and E-ink display ($\sim$220 ms) & 4.5 
   & 3.58 \\
optimized E-ink ($\sim$453 ms) / Unoptimized E-ink (2.8 s) & 5.5 / 5.21  & 5.32 / 5.13 \\
Idle
  between E-ink display and NBVLC TX/RX ($\sim$1.3 s) & 3.24
   & 3.2
  \\
NBVLC
  TX/RX ($\sim$81 ms) & 7.1
   & 6.8 \\
Idle
  after NBVLC TX/RX ($\sim$1.1 s) & 3.2
   & 3.1  \\
Deep
  sleep after E-ink display ($\sim$1:14.9 min) & 0.416
   & 0.344
  \\
{\textbf{Overall average current consumption during one cycle}} & \textbf{0.756 / 0.894} & \textbf{0.669 / 0.809}
\end{tblr}
\label{tab:verylowpower_nbvlc}
\vspace{-5mm}
\end{table}

\vspace{-3mm}
\section{Energy Modeling}
\label{sec:energy_models}
In this section, we present comprehensive models to predict the energy consumption of the Si-based RIoT device over its operational period, incorporating all phases from initial BLE communication (advertising and connected modes) and firmware startup sequence to cyclic sensing, E-ink display updates, and data transmission tasks. The energy consumption is calculated by considering the current draw and duration of each phase, thereby providing a detailed breakdown of energy usage across all activities. 
We developed three predictive energy models for the following Si-based RIoT node operations:
\begin{itemize}
    \item Normal (unoptimized) node operation
    \item Low-power (software-optimized) node operation
    \item Very low-power (hardware-optimized) node operation
\end{itemize}

\vspace{-3mm}
\subsection{Normal (unoptimized) node operation}
\label{sec:normal_oper}
To get an insight on the energy consumption of the node, let's consider the following scenario for the node operation:

\begin{methods}
    \item The node is powered on and starts BLE advertising in fast mode (advertising interval=20 ms). We consider a default BLE TX power of 0 dBm both during BLE advertising and connected states. 
    \item The timeout for the fast BLE advertising is set to 30 s.
    \item No central device is connected to the node while it is advertising in fast mode.
    \item Therefore, the node switches to slow BLE advertising mode (advertising interval=152.5 ms)
    \item A central device connects to the node after 30 s from the time the node started slow BLE advertising, with a default BLE connection interval of 45 ms.
    \item Then the node remains in BLE connected-only mode with the central device for another 30 s (i.e. no other operations are performed during this time).
    \item Once a central device connects to the node via BLE, the firmware test is initiated. A startup operation occurs (only once provided that the node does not go into a deep sleep state during its operation), followed by an idle period (see Fig. \ref{fig:startup}). The node then enters its repeated operational cycle: sensing, E-ink displaying, and NBVLC TX/RX, each followed by an idle period (see Fig. \ref{fig:unopt}).     
\end{methods}

Table \ref{tab:param} presents the default parameters used to calculate the node’s energy consumption.
These parameters were derived from comprehensive measurements taken across various operational states of the node. The energy model for this scenario can be formulated as follows:

\begin{figure}[t]
    \centering
    \subfigure[]{\includegraphics[width=0.65\linewidth]{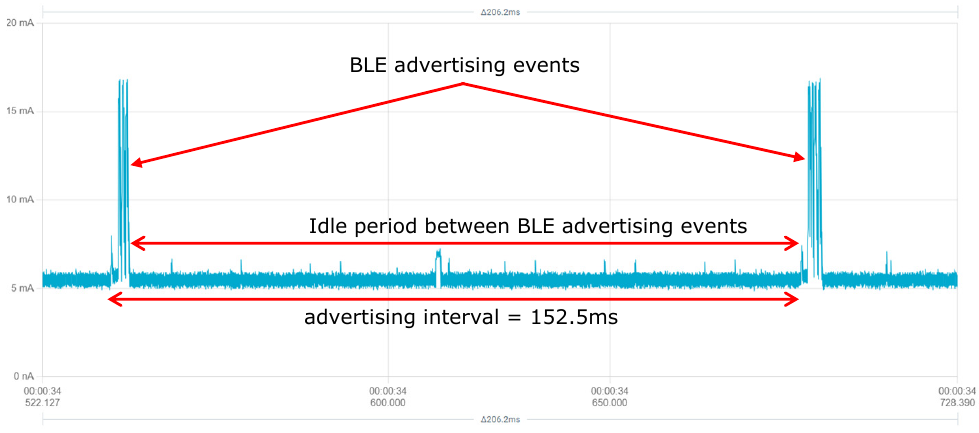}\label{fig:adv_int}}
    \hspace{0pt}
    \subfigure[]{\includegraphics[width=0.65\linewidth]{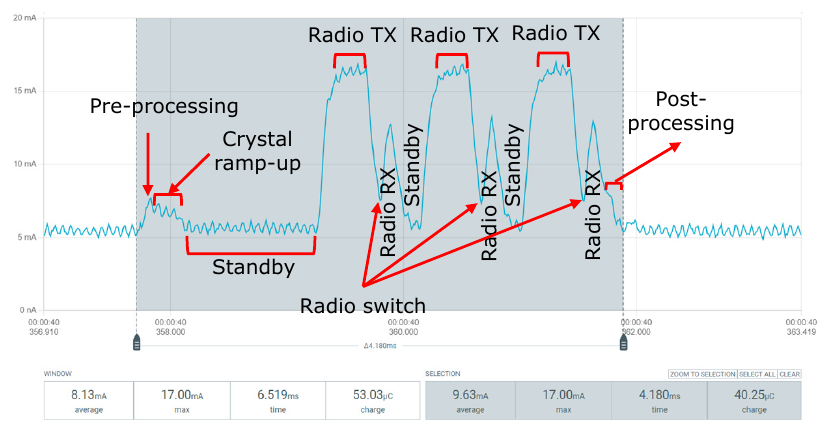}\label{fig:adv_event}}
    \caption{Current profile of (a) BLE advertising interval, and (b) BLE advertising event with different states.}
    \label{fig:ble_adv}
    \vspace{-7mm}
\end{figure}

\begin{methods}
\newmethod{\textbf{Charge consumption during BLE advertising intervals (fast and slow)}}: 

Referring to Fig. \ref{fig:ble_adv}, the average current consumption for a BLE advertising interval (fast or slow) can be written as follows:
\setlength{\belowdisplayskip}{2pt} \setlength{\belowdisplayshortskip}{2pt}
\setlength{\abovedisplayskip}{0pt} \setlength{\abovedisplayshortskip}{0pt}
\begin{equation}
I_{\text{advInt}}= \frac{( I_{\text{advEvent}} \times T_{\text{advEvent}}) + (I_{\text{advIdle}} \times T_{\text{advIdle}} )}{T_{\text{advEvent}} + T_{\text{advIdle}} }  \label{advint}
\end{equation}    
where $I_{\text{advEvent}}$ is the average current consumed during a BLE advertising event, $T_{\text{advEvent}}$ is the average duration of a single fast  advertising event, $I_{\text{advIdle}}$ is the average current consumed during the idle portion between two successive  advertising events, and $T_{\text{advIdle}}$ is the average duration of the idle period between two successive advertising events ($T_{\text{advIdle}}=T_{\text{slowAdvIdle}}$ for slow advertising mode or $T_{\text{advIdle}}=T_{\text{fastAdvIdle}}$ for fast advertising mode).
Therefore, using Eq. (\ref{advint}) and Table \ref{tab:param}, the average current consumption during fast and slow advertising intervals are $I_{\text{fastAdvInt}}=6.336 \ \text{mA}$ and $I_{\text{slowAdvInt}}=5.575 \ \text{mA}$, respectively.
Table \ref{tab:ble_adv_txpower} shows the average current consumption of the BLE advertising event for different TX power levels ($P_\text{tx}$), where it can be observed that the overall average current consumption for the node varies across different TX power levels, ranging from 
5.58 mA (at 0 dBm) to 5.75 mA (at +8 dBm).
During BLE advertising events when the device is actively broadcasting its presence (see Fig. \ref{fig:adv_event}), higher transmit power levels result in higher average current consumption. This is evident from the values ranging from 
9.65 mA (at 0 dBm) to 15.78 mA (at +8 dBm).
Idle current consumption between advertising events remains relatively constant across different TX power levels, as expected. 
The total charge consumption (in mAh) for fast advertising intervals is given by:
\begin{equation}
Q_{\text{fastAdvInt}} = I_{\text{fastAdvInt}}  \times T_{\text{fastAdvInt}} \times N_{\text{fastAdv}} ,
\label{q_fast_int}
\end{equation}  
where $T_{\text{fastAdvInt}}$ is the duration of a fast BLE advertising interval (in hours), and $N_{\text{fastAdv}}$ is the number of fast BLE advertising intervals over a given period of time, which can be computed as
\begin{equation}
N_{\text{fastAdv}} = \frac{T_{\text{fastAdv}}}{T_{\text{fastAdvInt}}},
\label{N_fast_int}
\end{equation}  
where $T_{\text{fastAdv}}$ is the duration for which the node advertises in fast mode during the initial operations.  
Similarly, the total charge consumption (in mAh) for slow advertising intervals is given by:
\begin{equation}
\begin{aligned}[t]
& Q_{\text{slowAdvInt}} = & I_{\text{slowAdvInt}}  \times T_{\text{slowAdvInt}} \times N_{\text{slowAdv}},
\end{aligned}
\label{q_slow_int}
\end{equation}
where $T_{\text{slowAdvInt}}$ is the duration of a slow BLE advertising interval (in hours), and $N_{\text{slowAdv}}$ is the number of slow BLE advertising intervals over a given period of time, which can be computed as
\begin{equation}
N_{\text{slowAdv}} = \frac{T_{\text{slowAdv}}}{T_{\text{slowAdvInt}}},
\label{N_slow_int}
\end{equation}  
where $T_{\text{slowAdv}}$ is the duration for which the node advertises in slow mode during the initial operations.

\begin{figure}
    \centering
    \includegraphics[width=0.65\linewidth]{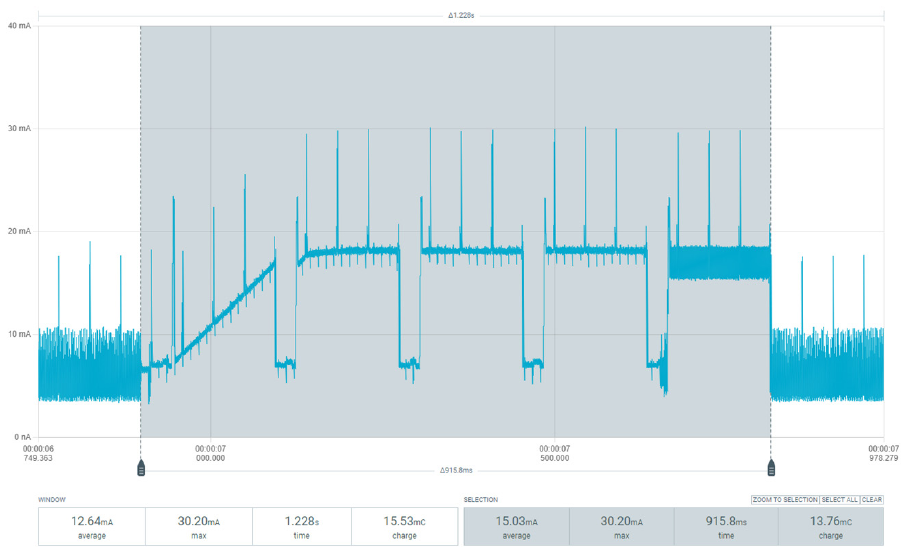}
    \caption{Illustration of startup current before the node starts its cyclic operations.
    }
    \label{fig:startup}
    \vspace{-7mm}
\end{figure}

\begin{figure}[t]
    \centering
    \subfigure[]
    {\includegraphics[width=0.65\linewidth]{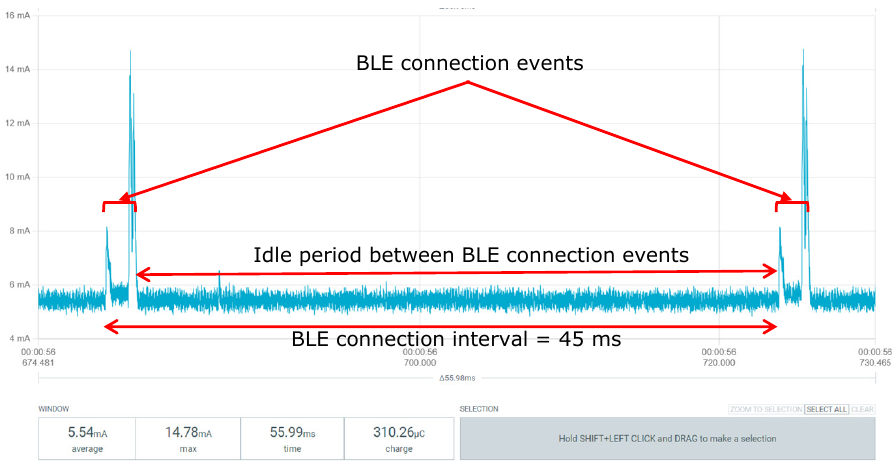}\label{fig:conn_int}}
    \hspace{0pt}
    \subfigure[]{\includegraphics[width=0.65\linewidth]{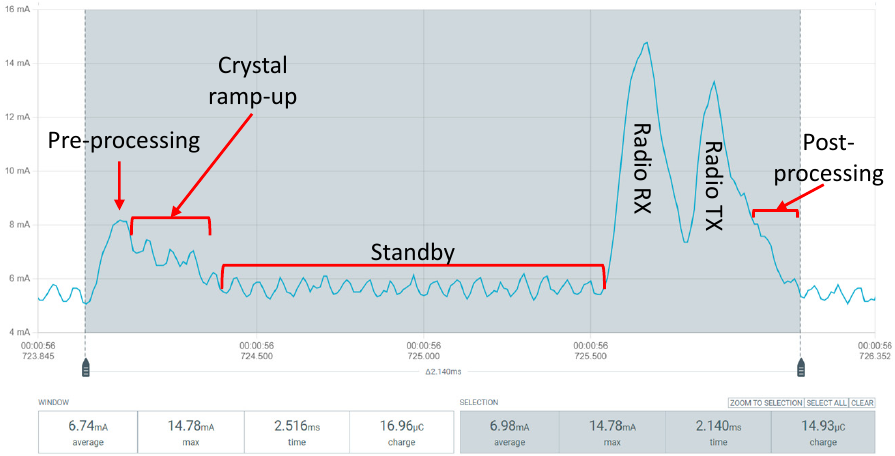}\label{fig:conn_event}}
    \caption{Current profile of (a) BLE connection interval, and (b) BLE connection event with different states.}
    \label{fig:ble_conn}
    \vspace{-7mm}
\end{figure}

\begin{table}[t]
\centering
\caption{Default parameters used to compute the energy of Si-based node during normal (unoptimized) and low power (software optimized) node operations.}
\label{tab:param}
\begin{tblr}{
  width = \linewidth,
  colspec = {Q[220]Q[360]Q[360]},
  row{1} = {c},
  cell{2}{2} = {c=2}{0.708\linewidth,c},
  cell{3}{2} = {c=2}{0.708\linewidth,c},
  cell{4}{2} = {c=2}{0.708\linewidth,c},
  cell{5}{2} = {c=2}{0.708\linewidth,c},
  cell{6}{2} = {c=2}{0.708\linewidth,c},
  cell{7}{2} = {c=2}{0.708\linewidth,c},
  cell{8}{2} = {c=2}{0.708\linewidth,c},
  cell{9}{2} = {c=2}{0.708\linewidth,c},
  cell{10}{2} = {c=2}{0.708\linewidth,c},
  cell{11}{2} = {c=2}{0.708\linewidth,c},
  cell{12}{2} = {c=2}{0.708\linewidth,c},
  cell{13}{2} = {c=2}{0.708\linewidth,c},
  cell{14}{2} = {c=2}{0.708\linewidth,c},
  cell{15}{2} = {c=2}{0.708\linewidth,c},
  cell{16}{2} = {c=2}{0.708\linewidth,c},
  cell{18}{2} = {c=2}{0.708\linewidth,c},
  cell{22}{2} = {c=2}{0.708\linewidth,c},
  cell{24}{2} = {c=2}{0.708\linewidth,c},
  cell{26}{2} = {c=2}{0.708\linewidth,c},
  cell{30}{2} = {c=2}{0.708\linewidth,c},
  cell{34}{2} = {c=2}{0.708\linewidth,c},
  hlines,
  vlines,
}
\textbf{Parameter} & \textbf{Normal (unoptimized)} & \textbf{Low-power (software-optimized) }\\
$V_{\text{operating}}$ & 3.3 V & \\
$P_\text{tx}$ & 0 dBm & \\
$T$ & 1 hour~ & \\
$T_\text{fastAdv}$ & 30 s & \\
$T_\text{slowAdv}$ & 30 s & \\
$T_\text{bleConn}$ & 30 s & \\
$T_\text{slowAdvInt}$ & 152.5 ms & \\
$T_\text{fastAdvInt}$ & 20 ms & \\
$T_\text{advEvent}$~ & 4.18 ms & \\
$T_\text{slowAdvIdle}$ & 148.32 ms & \\
$T_\text{fastAdvIdle}$ & 15.82 ms & \\
$I_\text{advIdle}$ & 5.46 mA & \\
$I_\text{advEvent}$ & 9.65 mA & \\
$T_\text{connInt}$ & 45 ms & \\
$T_\text{connEvent}$ & 2.14 ms & \\
$I_\text{connEvent}$ & 7.31 mA & 3.52 mA\\
$T_\text{connIdle}$ & 42.86 ms~ & \\
$I_\text{connIdle}$ & 5.43 mA & 1.33 mA\\
$I_\text{start}$ ~ ~~ & 15.03 mA & 14.3 mA\\
$T_\text{start}$ & 915.8 ms & 911.1 ms\\
$T_\text{idleStart}$ & 1.9 s \\
$I_\text{idleStart}$ & $I_\text{connInt}$=5.52 mA & $I_\text{connInt}$=1.434 mA\\
$T_\text{sens}$ & 516 ms & \\
$I_\text{sens}$ & 12.26 mA & 8.23 mA\\
$T_\text{idleSens}$ & 200 ms & \\
$I_\text{idleSens}$ & $I_\text{connInt}$=5.52 mA & $I_\text{connInt}$=1.434 mA\\
$T_\text{eink}$ & 2.8 s (unoptimized E-ink) & 0.435 s (optimized E-ink)\\
$I_\text{eink}$ & 6.84 mA & 3.63 mA\\
$T_\text{idleEink}$ & 2.1 s & \\
$I_\text{idleEink}$ & $1.007 \times I_\text{connInt}$=5.56 mA & $1.14 \times I_\text{connInt}$=1.635 mA\\
$T_\text{nbvlc}$ & 84 ms & 100 ms\\
$I_\text{nbvlc}$ & 8.33 mA & 8.15 mA\\
$T_\text{idleNbvlc}$ & 5.12 s & \\
$I_\text{idleNbvlc}$ & $1.007 \times I_\text{connInt}$=5.56 mA & $1.14 \times I_\text{connInt}$=1.635 mA
\end{tblr}
\vspace{-5mm}
\end{table}

\begin{table}[t]
\centering
\caption{Current draw as measured across node for different TX power levels during the BLE advertising state (BLE advertising interval = 152.5 ms).}
\begin{tblr}{
  width = \linewidth,
  colspec = {Q[160]Q[170]Q[270]Q[270]},
  hlines,
  vlines,
}
\textbf{$P_\text{tx}$ (dBm)} & \textbf{$I_{\text{advInt}}$ (mA) } & \textbf{$I_{\text{advIdle}}$ (mA)} $T_{\text{advIdle}}$=148.32 ms & \textbf{$I_{\text{advEvent}}$ (mA) }$T_{\text{advEvent}}$=4.18 ms\\
+8
   & 5.75 & 5.48 & 15.78\\
+4
   & 5.66 & 5.48 & 12.58\\
0
   & 5.58 & 5.46 & 9.65\\
\end{tblr}
\label{tab:ble_adv_txpower}
\vspace{-3mm}
\end{table}

\begin{table}
\centering
\caption{Current draw as measured across node for different BLE TX power levels during the normal and low-power BLE connected-only state (connection interval=45 ms).}
\label{tab:ble_conn_txpower}
\begin{tblr}{
  width = \linewidth,
  colspec = {Q[90]Q[173]Q[301]Q[301]},
  cell{1}{2} = {c=3}{0.821\linewidth,c},
  hlines,
  vlines,
}
 & \textbf{Normal (unoptimized) / Low-power (software-optimized)} &  & \\
\textbf{$P_\text{tx}$ (dBm)}  & \textbf{$I_{\text{connInt}}$ (mA)} & \textbf{$I_{\text{connIdle}}$ (mA)} $T_\text{connIdle}$=42.86 ms & \textbf{$I_{\text{connEvent}}$ (mA)} $T_\text{connEvent}$=2.14 ms\\
+8   & 5.59 / 1.48 & 5.44 / 1.33 & 8.58 / 4.69\\
+4   & 5.56 / 1.46 & 5.43~/ 1.33 & 8.12 / 4.20\\
0   & 5.51 / 1.43 & 5.43~/ 1.33 & 7.31 / 3.52
\end{tblr}
\vspace{-5mm}
\end{table}

\begin{table}[t]
\centering
\caption{Current draw as measured across node for different BLE connection intervals and BLE TX power levels when normal (unoptimized) sensing, and E-ink displaying operations are performed in a cycle.}
\begin{tblr}{
  width = \linewidth,
  colspec = {Q[105]Q[60]Q[80]Q[280]Q[115]Q[170]},
  cell{1}{3} = {c=4}{0.639\linewidth,c},
  cell{3}{1} = {r=3}{},
  cell{6}{1} = {r=3}{},
  cell{9}{1} = {r=3}{},
  cell{12}{1} = {r=3}{},
  vlines,
  hline{1-3,6,9,12,15} = {-}{},
  hline{4-5,7-8,10-11,13-14} = {2-6}{},
}
 &  & \textbf{Average current consumption (mA)} &  &  & \\
\textbf{$T_\text{connInt}$ (ms)} & \textbf{$P_\text{tx}$} & \textbf{\ Overall} & \textbf{$I_\text{idleEink}$ / $I_\text{idleSens}$}\quad $T_\text{idleEink}$=7.5 s / $T_\text{idleSens}$=200 ms & \textbf{$I_\text{sens}$}\quad $T_\text{sens}$= 516 ms  & \textbf{$I_\text{eink}$ (unoptimized)} $T_\text{eink}$=2.8 s\\
11.25~ & +8~ & 6.68~ & 6.03 / 5.95~ & 12.75~ & 7.30~\\
 & +4~ & 6.54~ & 5.91 / 5.89~ & 12.65~ & 7.21~\\
 & 0~ & 6.47~ & 5.84 / 5.79~ & 12.55~ & 7.12~\\
{
  45~\\(default)
    } & +8~ & 6.26~ & 5.62 / 5.58~ & 12.30~ & 6.91~\\
 & +4~ & 6.22~ & 5.59 / 5.56~ & 12.28~ & 6.88~\\
 & 0~ & 6.18~ & 5.56 / 5.50~ & 12.26~ & 6.84~\\
250~ & +8~ & 6.15~ & 5.53 / 5.47~ & 12.20~ & 6.81~\\
 & +4~ & 6.14~ & 5.53 / 5.47~ & 12.18~ & 6.79~\\
 & 0~ & 6.13~ & 5.50 / 5.47~ & 12.17~ & 6.78~\\
1000~ & +8~ & 6.12~ & 5.50 / 5.45~ & 12.13~ & 6.76~\\
 & +4~ & 6.12~ & 5.50 / 5.45~ & 12.12~ & 6.76~\\
 & 0~ & 6.11~ & 5.49 / 5.45~ & 12.09~ & 6.75~
\end{tblr}
\label{tab:normal_firmware01}
\vspace{-9mm}
\end{table}

\newmethod{ \textbf{Charge consumption during BLE connection intervals}:} \\
Referring to Fig. \ref{fig:ble_conn}, the average current consumption during a BLE connection interval is obtained as:
\begin{equation}
\begin{aligned}[t]
I_{\text{connInt}} = 
\frac{( I_{\text{connEvent}} \times T_{\text{connEvent}}) + 
(I_{\text{connIdle}} \times T_{\text{connIdle}} )}{T_{\text{connEvent}} + T_{\text{connIdle}} }  
\end{aligned}
\label{Ibleint}
\end{equation}
where $T_{\text{connEvent}}$ is the average duration of a single BLE connection event, $I_{\text{connEvent}}$ is the average current consumed during a BLE connection event, $T_{\text{connIdle}}$ is the idle period between two successive connection events, and $I_{\text{connIdle}}$ is the average current consumed during the idle period between two successive BLE connection events. 
Using Eq. (\ref{Ibleint}) and Table \ref{tab:param}, $I_{\text{connInt}}=5.52 \ \text{mA}$.
Table \ref{tab:ble_conn_txpower} 
also shows the average current consumption of the BLE connection interval for different TX power levels ($P_\text{tx}$). 
There is a noticeable decrease in the average current during BLE connection events with decreasing BLE TX power level from +8 dBm to 0 dBm (from 8.58 mA to 7.31 mA for normal unoptimized BLE connected-only state and from 4.69 mA to 3.52 mA for low-power software optimized BLE connected-only state). This indicates that reducing the TX power significantly lowers the current consumption during the active BLE connection events. The idle current, $I_{\text{connIdle}}$, between connection events remains fairly stable showing that idle periods between connection events are not affected by changes in TX power levels, as expected. The overall average current during a BLE connection interval, $I_{\text{connInt}}$  decreases slightly with reduced TX power levels, but the change is more modest compared to the BLE connection event current, $I_{\text{connEvent}}$.
The total charge consumption (in mAh) for BLE connection intervals is given by:
\begin{equation}
\begin{aligned}[t]
& Q_{\text{connInt}} =  & I_{\text{connInt}}  \times T_{\text{connInt}} \times N_{\text{connInt}},
\end{aligned}
\label{q_bleconn_int}
\end{equation}
where $T_{\text{connInt}}$ is the duration of a BLE connection interval (in hours), and $N_{\text{connInt}}$ 
is the number of instances of the BLE connection intervals over a given period of time during the initial operations, and it can be computed as:
\begin{equation}
N_{\text{connInt}} = \frac{T_{\text{bleConn}}}{T_{\text{connInt}}},  
\label{num_ble_conn}
\end{equation}  
where $T_{\text{bleConn}}$ is the time spent in BLE-connected only mode during the initial operations.

\newmethod{ \textbf{Charge consumption during startup}:} \\
As mentioned previously, the startup operation occurs only once when a firmware test mode is first activated (as illustrated in Fig. \ref{fig:startup}). The charge consumption (in mAh) during the startup activity is calculated as follows:
\begin{equation}
Q_{\text{start}} = I_{\text{start}}  \times T_{\text{start}}  ,
\label{q_startup}
\end{equation}  
where $T_{\text{start}}$ is the 
startup duration in hours, and $I_{\text{start}}$ is the average current consumed during the startup period (in mA).

\newmethod{ \textbf{Charge consumption during idle period after startup}: }\\
The charge consumption (in mAh) for the idle period that follows the startup operation can be computed as follows:
\begin{equation}
Q_{\text{idleStart}} = I_{\text{idleStart}}  \times T_{\text{idleStart}} ,
\label{q_idlestartup}
\end{equation}  
where $T_{\text{idleStart}}$ is the average duration of the idle period after startup (in hours, h), and $I_{\text{idleStart}}$ is the average current consumed during the idle period following startup (in mA).
Since the node is connected via BLE to a central device over the whole operating time, $I_{\text{idleStart}}$ corresponds to the average current consumption during a BLE connection interval, $I_{\text{connInt}}$, as computed in Eq. (\ref{Ibleint}).

\newmethod{ \textbf{Charge consumption during sensing operations}:} \\
The total charge consumed during the sensing operations (in mAh) can be formulated as:
\begin{equation}
Q_{\text{sens}} = I_{\text{sens}}  \times T_{\text{sens}} \times  N_{\text{sens}},
\label{q_sen}
\end{equation}  
where $T_{\text{sens}}$ is the average duration of a single sensing operation (in hours), $I_{\text{sens}}$ is the average current consumed during a sensing operation (in mA), and $N_{\text{sens}}$ is the number of instances of the sensing operations over the node’s operating duration. Table \ref{tab:normal_firmware01} shows the average current consumption during sensing under different BLE connection intervals ($T_\text{connInt}$) and TX power levels ($P_\text{tx}$).

\newmethod{ \textbf{Charge consumption during idle periods after sensing operations}:} \\
The total charge consumed during the idle periods after sensing (in mAh) can be expressed as:
\begin{equation}
\begin{aligned}[t]
& Q_{\text{idleSens}} =    & I_{\text{idleSens}}  \times T_{\text{idleSens}} \times  N_{\text{idleSens}},
\end{aligned}
\label{q_idlesensing}
\end{equation}  
where $T_{\text{idleSens}}$ is the average duration of the idle period after sensing (in hours), $I_{\text{idleSens}}$ is the average current consumed during the idle period following sensing (in mA),  and $N_{\text{idleSens}}$ is the number of instances of the idle period following the sensing operation over the node’s operating duration. Since the node is connected via BLE to a central device over the whole operating time, $I_{\text{idleSens}} \approx I_{\text{connInt}}$ (see Eq. (\ref{Ibleint}).

\newmethod{ \textbf{Charge consumption during E-ink display}: }\\
The total charge consumed during the E-ink displaying operations (in mAh) can be expressed as:
\begin{equation}
Q_{\text{eink}} = I_{\text{eink}}  \times T_{\text{eink}} \times  N_{\text{eink}},
\label{q_eink}
\end{equation}  
where $T_{\text{eink}}$ is the average duration of a single E-ink displaying operation (in hours), $I_{\text{eink}}$ is the average current consumed during the E-ink displaying operation (in mA), and $N_{\text{eink}}$ is the number of instances of the E-ink displaying operations over the node’s operating duration.
Table \ref{tab:normal_firmware01} shows the average current consumption during E-ink display under different BLE connection intervals ($T_\text{connInt}$) and TX power levels ($P_\text{tx}$).

\newmethod{ \textbf{Charge consumption during idle periods after E-ink display operations}:} \\
The total charge consumed during the idle periods after E-ink display operations (in mAh) can be written as:
\begin{equation}
\begin{aligned}[t]
& Q_{\text{idleEink}} =   & I_{\text{idleEink}}  \times T_{\text{idleEink}} \times  N_{\text{idleEink}},
\end{aligned}
\label{q_idle_eink}
\end{equation}  
where $T_{\text{idleEink}}$ is the average duration of the idle period after the E-ink display operation (in hours), $I_{\text{idleEink}}$ is the average current consumed during the idle period following the E-ink display operation (in mA), and $N_{\text{idleEink}}$ is the number of instances of the idle period following the E-ink display operations over the node’s operating duration.
Recall that the during the idle period after E-ink display, the node remains connected to a central device via BLE. 
Referring to Table \ref{tab:normal_firmware01} where we compared the impact of different TX power levels and BLE connection intervals on the node's average current consumption during sensing and E-ink displaying, we noticed an increase in the average idle current after E-ink displaying between 0.5\% and 0.9\% when compared to the average current during BLE connected-only mode  for the same BLE connection interval of 45 ms (see Table \ref{tab:ble_conn_txpower} for ``normal (unoptimized)" BLE connected-only state). Thus, the idle current after E-ink displaying, is considered to be on average, around 0.7\% higher than the BLE connected-only mode, i.e.
\begin{equation}
I_{\text{idleEink}} \approx  1.007 \times I_{\text{connInt}} = 5.56 \ \text{mA}. 
\end{equation}  
From Table \ref{tab:normal_firmware01}, we also note that with smaller connection intervals (e.g. 11.25 ms), the impact of increasing the BLE TX power level is greater on the overall average current consumption. However, with higher connection intervals such as 250 ms and 1000 ms, increasing the BLE TX power level has minimal impact on the overall current consumption.

\newmethod{ \textbf{Charge consumption during NBVLC TX/RX}:} \\
The total charge consumed during the NBVLC transmission and reception operations (in mAh) can be expressed as:
\begin{equation}
Q_{\text{nbvlc}} = I_{\text{nbvlc}}  \times T_{\text{nbvlc}} \times  N_{\text{nbvlc}},
\label{q_nbvlc}
\end{equation} 
where $T_{\text{nbvlc}}$  is the average duration of a single NBVLC transmission and reception operation (in hours), $I_{\text{nbvlc}}$ is the average current consumed during the NBVLC transmission and reception operation (in mA), and $N_{\text{nbvlc}}$ is the number of instances of the NBVLC TX/RX operations over the node’s operating duration. 	

\newmethod{ \textbf{Charge consumption during idle periods after NBVLC TX/RX operations}: }\\
The total charge consumed during the idle periods after the NBVLC TX/RX operations (in mAh) can be expressed as:
\begin{equation}
\begin{aligned}[t]
& Q_{\text{idleNbvlc}} = & I_{\text{idleNbvlc}}  \times T_{\text{idleNbvlc}} \times  N_{\text{idleNbvlc}},
\end{aligned}
\label{q_idle_nbvlc}
\end{equation} 
where $T_{\text{idleNbvlc}}$ is the average duration of the idle period after the NBVLC TX/RX operation (in hours), $I_{\text{idleNbvlc}}$ is the average current consumed during the idle period following a NBVLC TX/RX operation (in mA), and $N_{\text{idleNbvlc}}$ is the number of instances of the idle periods following the NBVLC TX/RX operation over the node’s operating duration.
We noted that the average idle current after NBVLC TX/RX followed the same trend as the average idle current after E-ink, when compared to the average current during BLE connected-only mode only. Thus, 
\begin{equation}
I_{\text{idleNbvlc}} \approx  1.007 \times I_{\text{connInt}} = 5.56 \ \text{mA},  
\end{equation}  
\end{methods}

\setlength{\belowdisplayskip}{2pt} \setlength{\belowdisplayshortskip}{2pt}
\setlength{\abovedisplayskip}{0pt} \setlength{\abovedisplayshortskip}{0pt}

The operations that are part of the duty cycle of the node and will be repeated over time consist of sensing, idle after sensing, E-ink displaying, idle after E-ink displaying, NBVLC TX/RX and idle after NBVLC TX/RX. Thus, the duration of a full cycle can be expressed as:
\begin{equation}
\begin{aligned}[t]
T_{\text{oneFullCycle}} = & \, T_{\text{sens}} + T_{\text{idleSens}} + T_{\text{eink}} + \\
& \, T_{\text{idleEink}} + T_{\text{nbvlc}} + T_{\text{idleNbvlc}}.
\end{aligned}
\label{duration_full_cycle}
\end{equation}
And the duration of the initial operations including BLE fast and slow advertising, BLE connected-only mode (idle), startup and idle after startup can be expressed as:
\begin{equation} 
\begin{aligned}[t]
T_{\text{initOper}} =  T_{\text{fastAdv}} + T_{\text{slowAdv}} + T_{\text{bleConn}} +  T_{\text{start}} + T_{\text{idleStart}} .
\end{aligned}
\label{duration_initial_oper}
\end{equation}
The number of cycles consisting of sensing, E-ink displaying and NBVLC operations (including the idle periods that follow these operations) over the node's operating time is obtained as follows:
\begin{equation}
\begin{aligned}[t]
N_{\text{cycles}} = \frac{(T - T_{\text{initOper}})}{T_{\text{oneFullCycle}}},
\end{aligned}
\label{num_sens_eink_nbvlc}
\end{equation}
where $T$ is the total operating time of the node.
The number of full cycles 
over the node's operating time can be expressed as:
\begin{equation}
\begin{aligned}[t]
N_{\text{fullCycles}} = \left\lfloor N_{\text{cycles}} \right\rfloor.
\end{aligned}
\label{eq:numcompletecycles}
\end{equation}
The total duration of the full cycles 
is then computed as:
\begin{equation}
\begin{aligned}[t]
T_{\text{fullCycles}} = &N_{\text{fullCycles}} \times  T_{\text{oneFullCycle}}.
\end{aligned}
\label{eq:duration_complete_cycles}
\end{equation}
In the event that there are both $N_{\text{fullCycles}}$
full cycles and also a partial cycle, we need to determine which operations will fit entirely within the partial cycle, which operations will only fit partially, and which operations will not fit at all. These need to be accounted for in the total charge consumption. 
The partial cycle duration 
can be computed as 
\begin{equation}
\begin{aligned}[t]
T_{\text{partCycle}} = (T - T_{\text{initOper}}) -  T_{\text{fullCycles}}.
\end{aligned}
\label{duration_partial_cycle}
\end{equation}
Next, we need to understand how the partial duration, $T_{\text{partCycle}}$, is distributed among the operations. Typically, operations occur sequentially within each cycle, so we proceed in order and determine how much of each operation fits into 
$T_{\text{partCycle}}$.
First, let’s write the total charge consumption for the partial cycle, $Q_{\text{partial}}$ (in simpler notations) as in Eq. (\ref{eq:partial}),  
where 
\( T_{\text{s}} \) and \( I_{\text{s}} \) denote the sensing duration ($T_{\text{sens}}$) and current ($I_{\text{sens}}$), respectively, 
\( T_{\text{i1}} \) and \( I_{\text{i1}} \) represent the idle period duration ($T_{\text{idleSens}}$) and current ($I_{\text{idleSens}}$) after sensing respectively (with \( I_{\text{i1}} \approx I_{\text{connInt}} \)), 
\( T_{\text{e}} \) and \( I_{\text{e}} \) correspond to the E-ink display duration ($T_{\text{eink}}$) and current ($I_{\text{eink}}$), respectively,
\( T_{\text{i2}} \) and \( I_{\text{i2}} \) refer to the idle period duration ($T_{\text{idleEink}}$) and current ($I_{\text{idleEink}}$) after the E-ink display respectively (with \( I_{\text{i2}} \approx 1.007 \times I_{\text{connInt}} \)), 
\( T_{\text{n}} \) and \( I_{\text{n}} \) are the NBVLC TX/RX duration ($T_{\text{nbvlc}}$) and current ($I_{\text{nbvlc}}$), respectively, and 
\( I_{\text{i3}} \) is the idle current ($I_{\text{idleNbvlc}}$) after NBVLC TX/RX (with \( I_{\text{i3}} \approx 1.007 \times I_{\text{connInt}} \)).

\vspace{-10mm}
\begin{strip}
\small
\begin{equation}
Q_{\text{partial}} = 
\left\{
\begin{aligned}
  & I_{\text{s}} \cdot T_{\text{partCycle}}, \quad  \text{if } T_{\text{partCycle}} \leq T_{\text{s}} \\ 
  & I_{\text{s}} \cdot T_{\text{s}} + I_{\text{i1}} \cdot (T_{\text{partCycle}} - T_{\text{s}}), \quad  \text{if } T_{\text{partCycle}} > T_{\text{s}} \text{ and } T_{\text{partCycle}} \leq T_{\text{s}} + T_{\text{i1}} \\ 
  & I_{\text{s}} \cdot T_{\text{s}} + I_{\text{i1}} \cdot T_{\text{i1}} + I_{\text{e}} \cdot (T_{\text{partCycle}} - T_{\text{s}} - T_{\text{i1}}), \quad  \text{if } T_{\text{partCycle}} > T_{\text{s}} + T_{\text{i1}} \text{ and } T_{\text{partCycle}} \leq T_{\text{s}} + T_{\text{i1}} + T_{\text{e}} \\ 
  & I_{\text{s}} \cdot T_{\text{s}} + I_{\text{i1}} \cdot T_{\text{i1}} + I_{\text{e}} \cdot T_{\text{e}} + I_{\text{i2}} \cdot (T_{\text{partCycle}} - T_{\text{s}} - T_{\text{i1}} - T_{\text{e}}), 
  \quad  \text{if } T_{\text{partCycle}} > T_{\text{s}} + T_{\text{i1}} + T_{\text{e}} \text{ and } T_{\text{partCycle}} \leq T_{\text{s}} + T_{\text{i1}} + T_{\text{e}} + T_{\text{i2}} \\ 
  & I_{\text{s}} \cdot T_{\text{s}} + I_{\text{i1}} \cdot T_{\text{i1}} + I_{\text{e}} \cdot T_{\text{e}} + I_{\text{i2}} \cdot T_{\text{i2}} + I_{\text{n}} \cdot (T_{\text{partCycle}} - T_{\text{s}} - T_{\text{i1}} - T_{\text{e}} - T_{\text{i2}}), 
  \quad  \text{if } T_{\text{partCycle}} > T_{\text{s}} + T_{\text{i1}} + T_{\text{e}} + T_{\text{i2}} \\ & \text{ and } T_{\text{partCycle}} \leq T_{\text{s}} + T_{\text{i1}} + T_{\text{e}} + T_{\text{i2}} + T_{\text{n}} \\ 
  & I_{\text{s}} \cdot T_{\text{s}} + I_{\text{i1}} \cdot T_{\text{i1}} + I_{\text{e}} \cdot T_{\text{e}} + I_{\text{i2}} \cdot T_{\text{i2}} + I_{\text{n}} \cdot T_{\text{n}} + I_{\text{i3}} \cdot (T_{\text{partCycle}} - T_{\text{s}} - T_{\text{i1}} - T_{\text{e}} - T_{\text{i2}} - T_{\text{n}}),  
  \\ & \text{if } T_{\text{partCycle}} > T_{\text{s}} + T_{\text{i1}} + T_{\text{e}} + T_{\text{i2}} + T_{\text{n}} 
\end{aligned}
\right.
\label{eq:partial}
\end{equation}
\vspace{-10mm}
\end{strip}

\noindent \textbf{Explanation of Eq. \ref{eq:partial}:}
\begin{methods}
    \item \textbf{Sensing:} If the remaining time $T_{\text{partCycle}}\leq T_{\text{s}}$, the sensing operation is only partially completed, and the charge consumption is $I_{\text{s}} \cdot T_{\text{partCycle}}$.
    
    \item \textbf{Idle after sensing:} If the remaining time $T_{\text{partCycle}} > T_{\text{s}}$ but less than or equal to $T_{\text{s}} + T_{\text{i1}}$, the sensing operation is fully completed, and part or all of the idle period after sensing is completed. The charge consumption accounts for the full sensing duration plus the idle time that fits.
    
    \item \textbf{E-ink:} 
        If $T_{\text{partCycle}}$ exceeds the combined time up to the E-Ink operation but is less than or equal to $T_{\text{s}} + T_{\text{i1}} + T_{\text{e}}$, it means that both the sensing and idle period after sensing fit fully, and part or all of the E-ink display operation fits.
    \item \textbf{Subsequent operations:} The pattern repeats for each operation, adding full contributions for the operations that fit entirely and partial contributions for the first operation that does not fit within the remaining time.
\end{methods}

The overall total charge consumption (in mAh) of the Si-based node during normal (unoptimized) operation can be expressed as:
\setlength{\belowdisplayskip}{1pt} \setlength{\belowdisplayshortskip}{1pt}
\setlength{\abovedisplayskip}{1pt} \setlength{\abovedisplayshortskip}{1pt}
\begin{equation}
\begin{aligned}[t]
 Q_{\text{total}} = & \, Q_{\text{fastAdvInt}} + Q_{\text{slowAdvInt}} + Q_{\text{startup}} + Q_{\text{sensing}} + \\ 
& \, Q_{\text{eink}} + Q_{\text{nbvlc}} + Q_{\text{idle}} + Q_{\text{partial}}
\end{aligned}
\label{q_all}
\end{equation}
where 
$Q_{\text{idle}}$ is the total charge consumption during the idle states in the initial operations (i.e BLE-connected only mode and idle after startup) and the idle periods in full operation cycles (i.e. idle states after sensing, E-ink displaying and NBVLC TX/RX), which can be expressed as:
\begin{equation}
\begin{aligned}[t]
   Q_{\text{idle}} &= Q_{\text{connInt}} + Q_{\text{idleStart}} + Q_{\text{idleSens}} +  Q_{\text{idleEink}} + Q_{\text{idleNbvlc}} \\ 
  & = I_{\text{connInt}} [ \left(T_{\text{connInt}} \times N_{\text{connInt}}\right) +  T_{\text{idleStart}} +  \\ & \quad  \left(T_{\text{idleSens}} \times N_{\text{fullCycles}}\right) +    \left(1.007 \times T_{\text{idleEink}} \times N_{\text{fullCycles}}\right) + \\ & \quad  \left(1.007 \times T_{\text{idleNbvlc}} \times N_{\text{fullCycles}}\right) ].
\end{aligned}
\label{q_idle}
\end{equation} 
Thus, we can rewrite Eq. \ref{q_all} as follows:
\begin{equation}
\begin{aligned}[t]
    Q_{\text{total}} 
    &= \left(I_{\text{fastAdvInt}} \times T_{\text{fastAdvInt}} \times N_{\text{fastAdv}}\right) + \\ 
    & \quad \left(I_{\text{slowAdvInt}} \times T_{\text{slowAdvInt}} \times N_{\text{slowAdv}}\right) +  
    \left(I_{\text{start}} \times T_{\text{start}}\right) + \\ 
    & \quad N_{\text{fullCycles}}  [ \left(I_{\text{sens}} \times T_{\text{sens}}\right) +  \left(I_{\text{eink}} \times T_{\text{eink}}\right) + \\ & \quad \left(I_{\text{nbvlc}} \times T_{\text{nbvlc}}\right)  ]   + 
     Q_{\text{idle}}
    +  Q_{\text{partial}}.
\end{aligned}
\label{eq:q_overall}
\end{equation}
The total charge consumption in Coulombs is given by:
\begin{equation}
Q_{\text{total}} \times 3.6,
\end{equation}
and the 
total energy consumption (in mWh) of the node 
over an operating duration of $T$ is given by:
\begin{equation}
P_{\text{total}} = Q_{\text{total}} \times V_{\text{operating}},
\end{equation}
where $V_{\text{operating}}$ is the operating voltage of the node.
Finally, the total energy consumption in Joules is given by:
\begin{equation}
P_{\text{total}} \times 3.6.
\end{equation}
The overall average current consumption (in mA) of the node 
over an operating duration of  $T$ is given by:
\begin{equation}
I_{\text{overall}} = \frac{
   I_{\text{1}} \times ( 
   T_{\text{initOper}} +  
   T_{\text{fullCycles}} 
 ) +  
   I_{2} \times T_{\text{partCycle}}
}
{\begin{aligned}
  & T
\end{aligned}}
\label{eq:I_overall}
\end{equation}
where $T= T_{\text{initOper}} + T_{\text{fullCycles}} + T_{\text{partCycle}} $, and
\begin{equation*}
I_{1} = \frac{\left( \begin{aligned}
  & Q_{\text{fastAdvInt}} + Q_{\text{slowAdvInt}} + Q_{\text{start}} + \\ 
  & Q_{\text{sens}} + Q_{\text{eink}} + Q_{\text{nbvlc}} + Q_{\text{idle}} 
\end{aligned} \right)}{
( 
  T_{\text{initOper}} + 
  T_{\text{fullCycles}} 
)},
I_{2} = \frac{Q_{\text{partial}}}{T_{\text{partCycle}}}.
\end{equation*}

Using the default parameters in Table \ref{tab:param} and equations (\ref{advint})-(\ref{eq:I_overall}), the overall energy consumption of the Si-based node in normal (unoptimized) operation mode is given in Table \ref{tab:energy_comp}.

\begin{table}[t]
\centering
\caption{Energy computation for normal, low-power and very low-power node operations (default parameters).}
\begin{tblr}{
  width = \linewidth,
  colspec = {Q[217]Q[191]Q[234]Q[279]},
  cell{1}{2} = {c=3}{0.704\linewidth,c},
  hlines,
  vlines,
}
 & \textbf{Node operation} &  & \\
\textbf{Computed energy parameters} & \textbf{Normal (unoptimized)} & \textbf{Low-power (software-optimized)} & \textbf{Very low-power (hardware-optimized)}\\
$Q_{\text{total}}$ & {6.22 mAh~\\(22.4 C)} & {2.27 mAh \\(8.17~C)} & {0.236 mAh~\\(0.85 C)}\\
$P_{\text{total}}$ & {20.53 mWh~\\(73.91 J)} & {7.49 mWh \\(26.96~J)} & {0.779 mWh~\\(2.81 J)}\\
$I_{\text{overall}}$ & 6.22 mA & 2.27 mA & 0.236 mA
\end{tblr}
\label{tab:energy_comp}
\vspace{-3mm}
\end{table}

\begin{table}[t]
\centering
\caption{Current draw as measured across node for different BLE connection intervals and BLE TX power levels when low power (software-optimized) firmware mode is activated (cyclic operations include sensing, and unoptimized E-ink displaying, including idle periods).}
\begin{tblr}{
  width = \linewidth,
  colspec = {Q[105]Q[60]Q[80]Q[280]Q[115]Q[170]},
  cell{1}{3} = {c=4}{0.656\linewidth,c},
  cell{3}{1} = {r=3}{},
  cell{6}{1} = {r=3}{},
  cell{9}{1} = {r=3}{},
  cell{12}{1} = {r=3}{},
  vlines,
  hline{1-3,6,9,12,15} = {-}{},
  hline{4-5,7-8,10-11,13-14} = {2-6}{},
}
 &  & \textbf{Average current consumption (mA)} &  &  & \\
\textbf{$T_\text{connInt}$ (ms)} & \textbf{$P_\text{tx}$} & \textbf{\ Overall} & \textbf{$I_\text{idleEink}$ / $I_\text{idleSens}$}\quad $T_\text{idleEink}$=7.5 s / $T_\text{idleSens}$=200 ms & \textbf{$I_\text{sens}$}\quad $T_\text{sens}$= 516 ms  & \textbf{$I_\text{eink}$ (unoptimized)} $T_\text{eink}$=2.8 s\\
11.25~ & +8~ & 2.80~ & 2.12 / 1.92~ & 8.76~ & 3.55~\\
 & +4~ & 2.72~ & 2.04 / 1.85~ & 8.64~ & 3.47~\\
 & 0~ & 2.58~ & 1.90 / 1.71~ & 8.52~ & 3.32~\\
45 \ \ (default) & +8~ & 2.35~ & 1.68 / 1.49~ & 8.28~ & 3.10~\\
 & +4~ & 2.33~ & 1.64 / 1.46~ & 8.26~ & 3.10~\\
 & 0~ & 2.31~ & 1.63 / 1.43~ & 8.23~ & 3.09~\\
250~ & +8~ & 2.23~ & 1.55 / 1.37~ & 8.17~ & 2.98~\\
 & +4~ & 2.22~ & 1.54 / 1.36~ & 8.15~ & 2.97~\\
 & 0~ & 2.21~ & 1.54 / 1.35~ & 8.14~ & 2.96~\\
1000~ & +8~ & 2.20~ & 1.53 / 1.35~ & 8.13~ & 2.95~\\
 & +4~ & 2.20~ & 1.53 / 1.33~ & 8.13~ & 2.95~\\
 & 0~ & 2.20~ & 1.53 / 1.33~ & 8.13~ & 2.95~
\end{tblr}
\label{tab:lowpowerfirmware}
\vspace{-5mm}
\end{table}

\begin{table*}
\centering
\caption{Validation of energy models.}
\label{tab:model_validation}
\begin{tblr}{
  width = \linewidth,
  colspec = {Q[45]Q[12]Q[12]Q[12]Q[12]Q[12]Q[12]Q[12]Q[12]Q[12]Q[12]Q[12]},
  cell{1}{2} = {c=4}{0.15\linewidth,c},
  cell{1}{6} = {c=4}{0.15\linewidth,c},
  cell{1}{10} = {c=3}{0.15\linewidth},
  cell{2}{2} = {c=2}{0.116\linewidth,c},
  cell{2}{4} = {c},
  cell{2}{5} = {c},
  cell{2}{6} = {c},
  cell{2}{7} = {c},
  cell{2}{8} = {c},
  cell{2}{10} = {c},
  cell{2}{11} = {c},
  cell{2}{12} = {c},
  cell{3}{2} = {c=2}{0.116\linewidth,c},
  cell{3}{4} = {c},
  cell{3}{5} = {c},
  cell{3}{6} = {c},
  cell{3}{7} = {c},
  cell{3}{8} = {c},
  cell{3}{10} = {c},
  cell{3}{11} = {c},
  cell{3}{12} = {c},
  cell{4}{2} = {c=2}{0.116\linewidth,c},
  cell{4}{4} = {c},
  cell{4}{5} = {c},
  cell{4}{6} = {c},
  cell{4}{7} = {c},
  cell{4}{8} = {c},
  cell{4}{10} = {c},
  cell{4}{11} = {c},
  cell{4}{12} = {c},
  cell{5}{2} = {c=2}{0.116\linewidth,c},
  cell{5}{4} = {c},
  cell{5}{5} = {c},
  cell{5}{6} = {c},
  cell{5}{7} = {c},
  cell{5}{8} = {c},
  cell{5}{10} = {c},
  cell{5}{11} = {c},
  cell{5}{12} = {c},
  cell{6}{2} = {c=2}{0.116\linewidth,c},
  cell{6}{4} = {c},
  cell{6}{5} = {c},
  cell{6}{6} = {c},
  cell{6}{7} = {c},
  cell{6}{8} = {c},
  cell{6}{10} = {c},
  cell{6}{11} = {c},
  cell{6}{12} = {c},
  cell{7}{10} = {c},
  cell{7}{11} = {c},
  cell{7}{12} = {c},
  cell{9}{10} = {c},
  cell{9}{11} = {c},
  cell{9}{12} = {c},
  cell{10}{10} = {c},
  cell{10}{11} = {c},
  cell{10}{12} = {c},
  cell{11}{10} = {c},
  cell{11}{11} = {c},
  cell{11}{12} = {c},
  cell{12}{10} = {c},
  cell{12}{11} = {c},
  cell{12}{12} = {c},
  cell{13}{10} = {c},
  cell{13}{11} = {c},
  cell{13}{12} = {c},
  cell{14}{10} = {c},
  cell{14}{11} = {c},
  cell{14}{12} = {c},
  cell{17}{5} = {c},
  cell{17}{7} = {c},
  cell{17}{9} = {c},
  cell{18}{4} = {c},
  cell{18}{9} = {c},
  cell{18}{11} = {c},
  cell{18}{12} = {c},
  cell{19}{2} = {c},
  cell{19}{3} = {c},
  cell{19}{4} = {c},
  cell{19}{5} = {c},
  cell{19}{6} = {c},
  cell{19}{7} = {c},
  cell{19}{8} = {c},
  cell{19}{9} = {c},
  cell{20}{2} = {c},
  cell{20}{3} = {c},
  cell{20}{4} = {c},
  cell{20}{5} = {c},
  cell{20}{6} = {c},
  cell{20}{7} = {c},
  cell{20}{8} = {c},
  cell{20}{9} = {c},
  cell{20}{10} = {c},
  cell{20}{11} = {c},
  cell{20}{12} = {c},
  hlines,
  vlines,
}
\textbf{Parameters} & \textbf{Normal unoptimized} &  &  &  & \textbf{Low power (software-optimized)} &  &  &  & \textbf{Very low power (hardware-optimized)} &  & \\
BLE & \ding{51} &  & \ding{51} & \ding{51} & \ding{51} & \ding{51} & \ding{51} & \ding{51} & \ding{55} & \ding{55} & \ding{55}\\
NBVLC & \ding{51} &  & \ding{55} & \ding{51} & \ding{51} & \ding{51} & \ding{51} & \ding{55} & \ding{51} & \ding{55} & \ding{55}\\
Sensing & \ding{51} &  & \ding{51} & \ding{51} & \ding{51} & \ding{51} & \ding{51} & \ding{51} & \ding{51} & \ding{51} & \ding{51}\\
E-ink display (Unoptimized) & \ding{51} &  & \ding{51} & \ding{55} & \ding{55} & \ding{55} & \ding{51} & \ding{55} & \ding{51} & \ding{55} & \ding{51}\\
E-ink display (optimized) & \ding{55} &  & \ding{55} & \ding{55} & \ding{51} & \ding{55} & \ding{55} & \ding{55} & \ding{55} & \ding{51} & \ding{55}\\
$P_\text{tx}$ (dBm) & 0 & 8 & 8 & 0 & 0 & 0 & 4 & 8 & - & - & -\\
$T$ (s) & 562 & 755 & 409.7 & 804 & 484.6 & 403.7 & 852 & 783 & 329.9 & 342.4 & 322.2\\
$T_\text{fastAdv}$ (s) & 18.92 & 25.59 & 27.91 & 18.09 & 23.67 & 26.61 & 25.93 & 17.81 & - & - & -\\
$T_\text{fastAdvInt}$ (ms) & 20 & 20 & 20 & 20 & 20 & 20 & 20 & 20 & - & - & -\\
$T_\text{slowAdv}$ (s) & 40.76 & 33.22 & 71.9 & 45.18 & 50.95 & 62.4 & 49.20 & 37.61 & - & - & -\\
$T_\text{slowAdvInt}$ (ms) & 152.5 & 152.5 & 152.5 & 152.5 & 152.5 & 152.5 & 152.5 & 152.5 & - & - & -\\
$T_\text{bleConn}$ (s) & 29.78 & 26.52 & 49.35 & 61.1 & 34.21 & 89.8 & 84.2 & 47.57 & - & - & -\\
$T_\text{connInt}$ (ms) & 45 & 11.25 & 11.25 & 45 & 45 & 45 & 250 & 45 & - & - & -\\
$T_\text{idleStart}$(s) & 7.02 & 7.4 & 7.542 & 7.02 & 6.8 & 7.02 & 7.02 & 7.3 & 0.03 & 0.03 & 0.03\\
$T_\text{idleSens}$ (ms) & 200 & 200 & 200 & 200 & 200 & 185 & 200 & 5400 & 220 & 118 & 118\\
$T_\text{idleEink}$ (s) & 2.1 & 2.07 & 7.642 & - & 1.9 & - & 2.1 & - & 1.3 & 0 & 0\\
$T_\text{idleNbvlc}$ (s) & 5.12 & 5.5 & - & 5.12 & 4.9 & 5.12 & 5.12 & - & 1.13 & - & -\\
$T_\text{deepSleep}$ (s) & - & - & - & - & - & - & - & - & 72.5 & 76.1 & 73.8\\
\textbf{U6 }cut / \textbf{U9 }cut & - & - & - & - & - & - & - & - & \ding{55} /\ding{55} & \ding{51} /\ding{51} & \ding{51} /\ding{51}\\
\textbf{Model prediction} (mWh) & \textbf{3.171} & \textbf{4.622} & \textbf{2.448} & \textbf{4.5} & \textbf{1.221} & \textbf{1.074} & \textbf{1.983} & \textbf{1.693} & \textbf{0.301} & \textbf{0.0827} & \textbf{0.108}\\
\textbf{Measured} (mWh) & \textbf{3.144} & \textbf{4.61} & \textbf{2.402} & \textbf{4.55} & \textbf{1.24} & \textbf{1.05} & \textbf{2.017} & \textbf{1.67} & \textbf{0.299} & \textbf{0.0821} & \textbf{0.111}\\
\textbf{Accuracy} & \textbf{99.1\%} & \textbf{\textbf{99.7\%}} & \textbf{98.1\%} & \textbf{98.9\%} & \textbf{98.4\%} & \textbf{97.7\%} & \textbf{98.3\%} & \textbf{98.6\%} & \textbf{99.3\%} & \textbf{99.3\%} & \textbf{97.3\%}
\end{tblr}
\vspace{-5mm}
\end{table*}

\vspace{-3mm}
\subsection{Low power (software-optimized) node operation}
\label{sec:lowpowermodel}
This section follows the same scenario, and operating principles outlined in Section \ref{sec:normal_oper}. That is, the node performs the same energy-intensive tasks — 
 sensing, E-ink displaying (in this case the optimized E-ink display is considered), and NBVLC transmission and reception (considering the same operating time, idle periods, BLE parameters, etc.) — while remaining connected to a central device via BLE throughout the entire operating period. 
Referring to the scenario described in Section \ref{sec:normal_oper}, the initial operations such as fast advertising and slow advertising  are performed in normal (unoptimized) mode. When a central device is connected to the node via BLE, the node will switch to a low-power (software-optimized) BLE connected-only mode (see Section \ref{software_opt}). 
Table \ref{tab:ble_conn_txpower} also shows the average current consumption during a BLE connection interval for different BLE TX power levels when the node is in low power (software-optimized) BLE connected-only mode. It can be observed from Table \ref{tab:ble_conn_txpower} that there is a modest increase (approx. 3.5\%) in the average current consumption ($I_{\text{connInt}}$) when the BLE TX power increases from 0 dB to +8 dBm (considering a BLE connection interval of 45 ms).
When the firmware test mode is activated, a startup operation will occur only once, and then the node will perform the cyclic operations, i.e. sensing, optimized E-ink displaying and NBVLC transmission and reception in low power mode (while the node is connected via BLE to a central device for the whole operating time).
Table \ref{tab:lowpowerfirmware} also shows the average current consumption measured across the node
for different BLE connection intervals and BLE TX power levels when low power (software-optimized) operation mode is activated and the node is performing sensing, (unoptimized) E-ink displaying and NBVLC TX/RX operations in a cycle (including idle periods).
From Table \ref{tab:lowpowerfirmware}, it can be observed that with smaller connection intervals (e.g. 11.25 ms), the impact of increasing the BLE TX power level is greater on the overall average current consumption. However, with higher connection intervals such as 250 ms and 1000 ms, increasing the BLE TX power level has minimal impact on the overall current consumption.

Similar to section \ref{sec:normal_oper}, 
$I_{\text{idleStart}}$ corresponds to the average current consumption during a BLE connection interval, $I_{\text{connInt}}$, which can be computed using Eq. (\ref{Ibleint}) and the parameters in Table \ref{tab:param} (see  column ``Low-power (software-optimized)").
The computed value for $I_{\text{connInt}}$ is also comparable to the measured value (see low power mode) obtained in Table \ref{tab:ble_conn_txpower} (considering 0 dBm TX power level and BLE connection interval of 45 ms).
From our measurements, we noted that while the 
idle periods after startup and sensing have comparable values to $I_{\text{connInt}}$, it is not the case for the idle period after E-ink displaying and also NBVLC TX/RX.
We noticed that the average current consumption of the idle state following the E-ink displaying and NBVLC TX/RX is approximately 14\% higher than $I_{\text{connInt}}$ (compare Table \ref{tab:ble_conn_txpower} with Table \ref{tab:lowpowerfirmware} for different BLE TX power levels, considering a BLE connection interval of 45 ms). Thus, for the low-power (software-optimized) node operation,
\setlength{\belowdisplayskip}{0pt} \setlength{\belowdisplayshortskip}{0pt}
\setlength{\abovedisplayskip}{0pt} \setlength{\abovedisplayshortskip}{0pt}
\begin{equation}
\begin{aligned}[t]
I_{\text{idleEink}}&=
I_{\text{idleNbvlc}} 
\approx  1.14 \times I_{\text{connInt}} =1.635 \ \text{mA}. 
\end{aligned}
\end{equation}  
Using the default parameters in Table \ref{tab:param}, the overall energy consumption of the Si-based node in low-power (software-optimized) operation mode is given in Table \ref{tab:energy_comp}.
Compared to the normal (unoptimized) node operation, we obtain a staggering $\sim$64\% decrease in energy consumption with the low-power (software-optimized) node operation, even though both scenarios involve the node performing the same tasks.

\vspace{-3mm}
\subsection{Very low power (hardware-optimized) node operation}
In this section, we consider the very low-power node operation scenario previously described in section \ref{hardware_opt}. The total charge consumption for the very low power node operation can be formulated as follows:
\begin{equation}
\begin{aligned}
  Q_{\text{total}} = & \, Q_{\text{start}} + Q_{\text{idleStart}} + Q_{\text{sens}} +  
  Q_{\text{idleSens}} + Q_{\text{eink}} + \\ 
  & \partial_{\text{nbvlc}} \times \left( Q_{\text{idleEink}} + Q_{\text{nbvlc}} + Q_{\text{idleNbvlc}} \right) + \\ 
  & Q_{\text{deepSleep}} + Q_{\text{partial}}  , 
\end{aligned}
\label{eq:q_total_verylowpower}
\end{equation}
where $\partial_{\text{nbvlc}}=1$ if the NBVLC functionality is enabled, otherwise $\partial_{\text{nbvlc}}=0$.
In the computations that follow, we consider the very low power firmware node operation, with the NBVLC functionality disabled ($\partial_{\text{nbvlc}}=0$), as well as BLE. We also consider the case where the node achieved the lowest current consumption, i.e., when both \textbf{U6} and \textbf{U9} measuring points were cut (see Table \ref{tab:verylowpower}).
Compared to sections \ref{sec:normal_oper} and \ref{sec:lowpowermodel}, the startup operation will be part of the operating cycle and will be repeated over the node’s operating duration, along with the sensing, optimized E-ink-displaying, and deep sleep operations (including any idle periods).
We compute the overall energy consumption of the node in the very low power operation mode using the parameters from Table \ref{tab:verylowpower}.
The duration of a full cycle consisting of startup, sensing, optimized E-ink displaying and deep sleep operations (including the idle periods) can be written as
\begin{equation}
\begin{aligned}  T_{\text{oneFullCycle}} = & \, T_{\text{start}} + T_{\text{idleStart}} +  T_{\text{sens}} +  
   T_{\text{idleSens}} + \\ & T_{\text{eink}} + T_{\text{deepSleep}}  , 
\end{aligned}
\end{equation}
where 
$T_{\text{deepSleep}}$ 
represent the time duration for the deep sleep state.
The number of full cycles over the node’s operating time is then given by
\begin{equation}
\begin{aligned}  
N_{\text{fullCycles}}  =\left\lfloor \frac{T}{T_{\text{oneFullCycle}}} \right\rfloor. 
\end{aligned}
\end{equation}
Thus, the total duration of the full cycles ($T_{\text{fullCycles}}$) is computed in the same way as in Eq. \ref{eq:duration_complete_cycles}.
Therefore, the amount of time available for a partial cycle can be expressed as
\begin{equation}
\begin{aligned}  
 T_{\text{partCycle}} = 
 T - T_{\text{fullCycles}} .
\end{aligned}
\end{equation}
The total charge consumption for the partial cycle, $Q_{\text{partial}}$ can be formulated in a similar way as in Eq. (\ref{eq:partial}).
Assuming $\partial_{\text{nbvlc}}=0$, Eq. (\ref{eq:q_total_verylowpower}) can be re-written as
\begin{equation} 
\begin{aligned} 
Q_{\text{total}} = & N_{\text{fullCycles}} \times [
\left( I_{\text{start}} \times T_{\text{start}} \right) +
\left( I_{\text{idleStart}} \times T_{\text{idleStart}} \right) + \\&
\left( I_{\text{sens}} \times T_{\text{sens}} \right) + \left( I_{\text{idleSens}} \times T_{\text{idleSens}} \right) + \left( I_{\text{eink}} \times T_{\text{eink}} \right) \\& + \left( I_{\text{deepSleep}} \times T_{\text{deepSleep}} \right) ] + Q_{\text{partial}}, 
\end{aligned} 
\end{equation}
where 
$I_{\text{deepSleep}}$ represent the average current consumption for the 
deep sleep state.
The overall average current consumption (in mA) of the node in very low power (hardware-optimized) operation mode over an operating duration of $T$ is given by:
\begin{equation} 
I_{\text{overall}} = \frac{ \left( I_{1} \times T_{\text{fullCycles}} \right) +
\left( I_{2} \times  T_{\text{partCycle}} \right) }{T} ,
\end{equation}
where $T=T_{\text{fullCycles}} + T_{\text{partCycle}}$, and
\begin{equation*}
I_{\text{1}} = \frac{
\left( Q_{\text{start}} + Q_{\text{idleStart}} + 
Q_{\text{sens}} + Q_{\text{idleSens}} + 
Q_{\text{eink}} + Q_{\text{deepSleep}} \right)
}{
T_{\text{fullCycles}}
}
\end{equation*}
and 
$I_{\text{2}} = \frac{
Q_{\text{partial}}
}{
T_{\text{partCycle}}
}.$

Using the average current consumption values in Table \ref{tab:verylowpower} (considering both \textbf{U6} and \textbf{U9} cut) and considering an operating time of $T$=1 hour, the overall energy consumption of the Si-based node in the very low power (hardware-optimized) operation mode is given in Table \ref{tab:energy_comp}. It can be concluded that disabling the communication modules while allowing the node to remain active for only a brief period before entering a prolonged deep sleep mode significantly reduces the overall energy consumption of the Si-based node.

\vspace{-3mm}
\section{ Model Validation}
The developed energy models for the Si-based node in section \ref{sec:energy_models} — covering (i) normal, unoptimized operation, (ii) low-power, software-optimized operation, and (iii) very low-power, hardware-optimized operation — were thoroughly validated using new measurement data obtained from the node under various configurations and scenarios, as shown in Table \ref{tab:model_validation}.
These validation tests ensured that the models accurately captured the energy consumption characteristics across different operating states. The models achieved an accuracy exceeding 97\%, indicating their reliability for predicting energy usage in diverse node configurations.

\vspace{-3mm}
\section{Conclusion}
The energy characterization of the SUPERIOT Si-based RIoT node has established a solid foundation for energy optimization and sustainability in the ongoing development of hybrid and fully printed IoT nodes. Through extensive measurements across various operating states, a comprehensive understanding of the node’s energy consumption was achieved. The insights gained from these measurements informed the development of models that accurately predict energy consumption under three different configurations, achieving a validation accuracy of over 97\%. 
This high level of accuracy will be instrumental for energy budgeting and optimization of future node iterations.
As the project advances to incorporate energy harvesting in the hybrid node and printed electronics in the final implementation (i.e. fully-printed node), these models will play a critical role in guiding energy management strategies. They will enable the design of nodes that operate efficiently under varying energy availability conditions, ensuring continuous functionality even in resource-constrained scenarios. Furthermore, the ability to benchmark the current Si-based node's energy performance against future node iterations will facilitate a structured approach to achieving energy sustainability, ensuring that each design iteration builds upon the efficiency gains of its predecessor.

\vspace{-3mm}
 \section*{Conflict of interest statement} 
The authors declare that they have no conflict of interest.

\vspace{-3mm}
\section*{Data availability statement}
No new data were created in this study.

\vspace{-3mm}
\section*{Acknowledgments}
The SUPERIOT project has received funding from the Smart Networks and Services Joint Undertaking (SNS JU) under the European Union’s Horizon Europe research and innovation programme under Grant Agreement No 101096021, including funding under the UK government’s Horizon Europe funding guarantee, UKRI Grant Reference Number 10053751. 
\vspace{-5mm}

\bibliographystyle{IEEEtran}
\bibliography{bibliography}
\end{document}